\documentclass{article}

\usepackage{latexsym}
\usepackage[blocks]{authblk}
\newcommand{\xh}{\textbf{x}}

\usepackage[english]{babel} 
\usepackage{graphicx}
\usepackage{amsmath}
\usepackage{amsfonts}
\usepackage{epstopdf}
\usepackage[colorlinks=true]{hyperref}
\hypersetup{
	linkcolor = blue,
	citecolor = red
}
\usepackage{xcolor}
\usepackage{color}
\usepackage{titlesec}
\makeatletter
\newif\if@secnum
\definecolor{lgray}{gray}{0.75}
\definecolor{blue1}{rgb}{0.0, 0.47, 0.85}
\definecolor{mDarkTeal}{HTML}{23373b}
\definecolor{white1}{rgb}{0.95, 0.95, 0.96}
\definecolor{mLightGreen}{HTML}{14B03D}
\definecolor{blue2}{rgb}{0.89, 0.94, 0.96}
\definecolor{darkseagreen}{rgb}{0.86, 0.95, 0.86}
\titleformat{\section}[hang]{\Large\bfseries\color{black}}{%
	\global\@secnumtrue
}{0em}{%
{%
	\setlength{\fboxsep}{0pt}%
	\colorbox{white}{\makebox[\textwidth]{\Large\strut}}%
}%
\hspace*{-\textwidth}%
\if@secnum%
\,\arabic{section}%
\hspace*{1em}%
\fi%
\global\@secnumfalse%
}[]
\makeatother
\newtheorem{remark}{Remark}[section]
\newtheorem{proposition}{Proposition}[section]

\usepackage{algorithm,algorithmicx}
\usepackage{algpseudocode}
\usepackage{mathdots}
\usepackage{graphicx} 
\usepackage{enumerate}
\usepackage{relsize}

\begin{document}

\title{Semi-explicit solutions to the water-wave dispersion relation and their role in the nonlinear Hamiltonian Coupled-Mode theory}
\author[1]{T.K. Papathanasiou }
\author[2]{Ch.E. Papoutsellis }  
\author[3]{G.A. Athanassoulis }
\affil[1]{ Department of Mechanical, Aerospace and Civil Engineering, Brunel University London, Uxbridge UB8 3PH, UK\\
	email: theodosios.papathanasiou@brunel.ac.uk}
\affil[2]{ \'{E}cole Centrale Marseille and Institut de Recherche sur les Ph\'{e}nom\`{e}nes Hors Equilibre (IRPHE), Marseille, France}
\affil[3]{ 	National Technical University of Athens, Zografos, Greece and
		Research Center for High Performance Computing, ITMO University, St. Petersburg, Russian Federation}
\date{}

\maketitle

\begin{abstract}
The Hamiltonian Coupled-Mode Theory (HCMT), recently derived by Athanassoulis and Papoutsellis \cite{Ref1}, provides an efficient new approach for solving fully nonlinear water-wave problems over arbitrary bathymetry. This theory exactly transforms the free-boundary problem to a fixed-boundary one, with space and time varying coefficients. In calculating these coefficients, heavy use is made of the roots of a local, water-wave dispersion relation with varying parameter, which have to be calculated at every horizontal position and every time instant. Thus, fast and accurate calculation of these roots, valid for all possible values of the varying parameter, are of fundamen-tal importance for the efficient implementation of HCMT. In this paper, new, semi-explicit and highly accurate root-finding formulae are derived, especially for the roots corresponding to evanescent modes. The derivation is based on the successive application of a Picard-type iteration and the Householder’s root finding method. Explicit approximate formulae of very good accuracy are obtained, and machine-accurate determination of the required roots is easily achieved by no more than three iterations, using the explicit forms as initial values. Exploiting this procedure in the HCMT, results in an efficient, dimensionally-reduced, numerical solver able to treat fully non-linear water waves over arbitrary bathymetry. Applications to four demanding nonlinear problems demonstrate the efficiency and the robustness of the present approach. Specifically, we consider the classical tests of strongly nonlinear steady wave propagation and the transformation of regular waves due to trapezoidal and sinusoidal bathymetry. Novel results are also given for the disintegration of a solitary wave due to an abrupt deepening. The derived root-finding formulae can be used with any other multimodal methods as well.


\end{abstract}

\tableofcontents

\section{Introduction}
\label{Intro}
The simulation of nonlinear water waves over variable bathymetry is a challenging task that involves various numerical techniques and extensive computations. The complications due to the presence of the unknown free-surface elevation and the varying bathymetry are usually treated either by perturbation techniques or by direct numerical methods (DNM). The first approach, being in use for more than a century, has led to a plethora of approximate models, mainly classified as Boussinesq-type or Serre-Green-Naghdi-type models \cite{Ref2}, \cite{Ref3}, \cite{Ref4}, \cite{Ref5}, \cite{Ref6}, \cite{Ref7}, \cite{Ref8}, \cite{Ref9}. DNM include finite-difference methods \cite{Ref10}, finite-element methods \cite{Ref11}, \cite{Ref12} and boundary-element (BEM) methods \cite{Ref13}, \cite{Ref14}. Perturbative approaches result in numerically efficient, dimensionally reduced models, limited to weakly nonlinear phenomena, while DNM are able to accurately treat strongly nonlinear/dispersive problems at the expense of efficiency, because of their high computational cost.

Recently, \cite{Ref1}, see also \cite{Ref15}, proposed a new formulation of the exact water-wave problem over arbitrary (smooth) bathymetry, providing both dimensional reduction and high accuracy, comparable with that ensured by DNM. This formulation is based on the exact semi-separation of variables in the instantaneous (non-canonical) fluid domain, established in \cite{Ref16}, referred subsequently as AP17. The wave potential $\Phi  = \Phi  ( \xh , z , t )$, where $\xh = ( x_{ 1}  ,  x_{ 2}  )$ is the horizontal position in the fluid domain, is expanded in a rapidly convergent series of the form $\Phi     =$ $\sum \varphi _{n} (  \xh  ,  t  )    Z_{n} \left({\kern 1pt} z  {\kern 1pt} ;  {\kern 1pt} \eta   ,  h  \right) $, where the local vertical functions $Z_{  n} $ are defined in the varying interval $\left[-  {\kern 1pt} h  (  \xh  )  ,  {\kern 1pt} \eta   (  \xh  ,  t  ){\kern 1pt} \right]$, delimited by the local depth $h  (  \xh  )$ and the local, instantaneous free-surface elevation $\eta   (  \xh  ,  t  )$. Introducing this expansion in Luke's variational principle, we obtain a convenient system of two nonlinear Hamiltonian evolution equations with respect to $\eta   (  {\bf x}  ,  {\kern 1pt} t  )$ and the surface potential $\psi   (  {\bf x}  ,  {\kern 1pt} t  )$ (see Eqs. \eqref{GrindEQ__7_}), containing a non-local coefficient field, accounting for the substrate kinematics. The latter is determined by a linear, coupled-mode system of horizontal differential equations with variable coefficients, dependent also on the unknown free-surface elevation (see Eqs. \eqref{GrindEQ__8_}). The last problem can be solved at each discrete time $t$ using the free-surface elevation prediction from the previous time step, providing a versatile and numerically efficient substitute for the Dirichlet-to-Neumann (DtN) operator, introduced by Craig \& Sulem \cite{Ref17}. This approach is called subsequently Hamiltonian Coupled-Mode Theory or System (HCMT or HCMS), according to the context. In comparison with the nonlinear consistent coupled-mode system, derived previously in \cite{Ref18}, the HCMS is more efficient and more accurate, since a double-series term appearing in the former has been summed analytically in the latter.

The approximations involved in the numerical implementation of the HCMS, apart from the inevitable discretization of the free-surface elevation $\eta   (  {\bf x}  ,  {\kern 1pt} t  )$ and the depth function $h  (  {\bf x}  )$, constitute of the truncation of the series expansion for the wave potential, and the calculation of the coefficients of the substrate system (see Eqs. \eqref{GrindEQ__8_} - \eqref{ABCs}). Since the series expansion is uniformly and rapidly convergent, even for strongly deformed boundaries (see AP17), the only critical issue is the accurate and fast calculation of the coefficients, which have to be calculated at any point of the spatial and temporal discretization. These coefficients are defined as integrals of the vertical basis functions, along local vertical intervals throughout the fluid.

The vertical basis functions are constructed by extending an $L^{2} -$basis to an $H^{2} -$basis (Sobolev space basis) as explained in AP17. The $L^{2} -$basis may be constructed by means of the eigenfunctions of any regular Sturm-Liouville problems, defined in the local vertical intervals $(-  h  ,  {\kern 1pt} \eta   ){\kern 1pt} $. Making the plausible choice to consider the Sturm-Liouville problem corresponding to linear water-wave problem in the strip $(-  h  ,  {\kern 1pt} \eta   ){\kern 1pt} $, results in the usual eigenfunctions, with eigenvalues defined through the dispersion relation of linear water waves with a frequency parameter varying in space and time. Since the coefficients of the substrate kinematical problem, Eq. \eqref{GrindEQ__8_}, are eventually expressed analytically in terms of these eigenvalues, the accurate and robust determination of the latter is of fundamental importance for the numerical implementation of the HCMS. Note that, in the computation of a demanding nonlinear problem we need to evaluate a number of 5 -- 8 eigenvalues for $10^{  8} $ (in 2D cases) up to $10^{  11} $ (in 3D cases) times. Thus, ideally, we would like to have explicit formulae or, at least, to ensure high accuracy with a few (1 -- 3) Newton-Raphson iterations. Even more essential is the robustness of the root finder, since in the case of non-convergence at any specific point-time the solution procedure will be stopped.

Attempts to determine explicit approximate formulae for the solution of the linear water-wave dispersion relation goes back to 50s, aiming -at that time- mainly to obtaining a convenient formula for the wavelength corresponding to a given frequency and depth. See, among others, \cite{Ref19}, \cite{Ref20}, \cite{Ref21}, \cite{Ref22}, \cite{Ref23}, \cite{Ref24}, \cite{Ref25}, \cite{Ref26}, \cite{Ref27}, \cite{Ref28}. More recently, several authors have focused on obtaining very accurate expressions for the real root, corresponding to the propagating mode. See e.g. \cite{Ref29}, \cite{Ref30}, \cite{Ref31}, \cite{Ref32}. An interesting observation is that, among the aforementioned plethora of studies, only few refer in detail to the roots associated with evanescent modes. These include basically the analysis of \cite{Ref23}, \cite{Ref25} and \cite{Ref27}, \cite{Ref28}. This fact is partially justified by the physical significance of the propagating mode, linked to the real root. However, several recent advances regarding the modelling and solution of challenging ocean and coastal engineering problems necessitate the very accurate and effective (in terms of computational time) determination of a number of evanescent modes as well. Apart from our HCMT, which is a fully nonlinear approach needing a huge number of accurate evaluation of several eigenvalues, various other formulations also rely on the evaluation of several eigenvalues as, for example, the extended mild-slope equations \cite{Ref33}, \cite{Ref34}, \cite{Ref35}, and the consistent coupled-mode systems \cite{Ref36}, \cite{Ref37}, \cite{Ref38}, \cite{Ref39}.

In this study, new explicit and semi-explicit (requiring 1 to 3 iterations), highly accurate formulae for the imaginary roots of the water-wave dispersion relation are presented and analysed. These formulae occur from the iteration of recursive numerical root-finding schemes, especially fitted for the nonlinear equation under consideration. The explicit ones provide an accuracy of $10^{  -5} $ or better for all eigenvalues and for all values of the varying parameter, and can be used directly in the solvers of linear and non-strongly nonlinear water-wave problems. The semi-explicit ones provide machine accuracy of $10^{  -15} $ with two or three iterations, and are suitable for demanding long-time simulations using the HCMS. This suitability is clearly demonstrated by numerical simulations implemented by means of HCMS and the above root-finding methods, in three nonlinear problems; the propagation of highly nonlinear travelling waves over flat bottom, the transformation of a regular wave by a submerged trapezoidal bar \cite{Ref40}, \cite{Ref41}, and the reflection of nonlinear waves due to a sinusoidal bottom patch \cite{Ref42}. After establishing its effectiveness and accuracy, the present method is utilized in studying the transformation (disintegration) of a solitary wave passing over an abrupt deepening. The phenomenon revealed in this case is studied herein for the first time, to the best of our knowledge.

The paper is organised as follows: in Section \ref{sec:2}, the HCMT for the evolution of nonlinear water-waves over arbitrary bathymetry is briefly presented, and its efficient numerical implementation is discussed in Section \ref{sec:3}. In Section \ref{sec:4}, the root approximation strategy is developed and analysed. Two iterative procedures (one of second and one of third order) are formulated and error estimates are derived. The procedures for obtaining explicit approximate formulae for the roots, and a thorough analysis of the performance of these formulae either as explicit approximations or as improved initializations of iteration schemes, are the subject of Section \ref{sec:5}. Section \ref{sec:6} is devoted to the investigation of the performance of the new formulae in the simulation of the three benchmark nonlinear water-wave problems mentioned above, and in the investigation of a new phenomenon occurring when a solitary wave pass over an abrupt deepening. In Section \ref{sec:7}, a general discussion is presented and main conclusions are summarized. Finally, in a technical appendix, the proof of a convergence lemma is given.

\section{Hamiltonian coupled-mode formalism for nonlinear water waves}
\label{sec:2}
In this section we briefly present a complete account of the new HCMT for fully nonlinear water waves over arbitrary bathymetry. To address the similarities and distinctions between the present theory and the well-developed, classical Hamiltonian theory by Zakharov \cite{Ref43} and Craig \& Sulem \cite{Ref17}, we start by a quick review of the latter.

\subsection{The classical Hamiltonian approach}
The fully nonlinear water-wave problem in the two horizontal dimensions admits of a Hamiltonian formulation in terms of the free surface elevation $\eta   (  \xh  ,  {\kern 1pt} t  )$ and the trace of the wave potential on the free surface, $\psi   (  \xh  ,  {\kern 1pt} t  )    =      \Phi   (  \xh  ,  {\kern 1pt} z=\eta   (  \xh  ,  {\kern 1pt} t  )  ,  {\kern 1pt} t  )$, as canonical variables \cite{Ref17}, \cite{Ref43}, \cite{Ref44}. In this context, the wave motion is governed by the two Hamiltonian evolution equations

\begin{subequations}\label{GrindEQ__1_}
	\begin{align}
	\partial_t\eta &= \mathcal{G}[\eta,h]\psi,\\
	\partial_{t}\psi &=-g\eta -\frac{1}{2}|\nabla_{\xh}\psi|^2+\frac{\big(\mathcal{G}[\eta,h]\psi+\nabla_{\xh}\psi\cdot\nabla_{\xh}\eta\big)^2}{2\left(1+
		|\nabla_{\mathbf{x}}\eta|^2\right)}.
	\end{align}
\end{subequations}

where ${\rm {\mathcal G}}  {\kern 1pt} [  \eta   ,  h  ]  {\kern 1pt} \psi $ denotes the Dirichlet-to-Neumann (DtN) operator defined by the formula

\begin{equation} \label{GrindEQ__2_}
{\rm {\mathcal G}} [ \eta  ,  h  ]  {\kern 1pt} \psi       =      -\nabla_{\mathbf{x}} \eta   \cdot {\kern 1pt}   \left[{\kern 1pt} \nabla_{\mathbf{x}}   \Phi   \right]_{{\kern 1pt} z  {\kern 1pt} ={\kern 1pt}   \eta }     +      \left[{\kern 1pt} \partial _{z}   \Phi   \right]_{{\kern 1pt} z  {\kern 1pt} ={\kern 1pt}   \eta } .
\end{equation}
In Eqs. \eqref{GrindEQ__1_}-\eqref{GrindEQ__2_}, $\nabla _{x}     =    (  \partial _{x_{1} }   ,    \partial _{x_{2} } )$ and $\nabla     =    (  \partial _{x_{1} }   ,    \partial _{x_{2} }   ,    \partial _{z}   )$ denote the horizontal and three-dimensional (3D) gradients, respectively. Further, the wave potential $\Phi     =    \Phi (  \xh  ,  z  ,  t  )$ satisfies the Laplace equation in the fluid domain together with the bottom impermeability condition on the seabed, and appropriate lateral conditions. Albeit a linear problem, the implementation of the non-local DtN operator, Eq. \eqref{GrindEQ__2_}, which takes care of the substrate (interior and bottom) kinematics of the fluid, is the most crucial step for demanding numerical simulations of nonlinear water-waves, in terms of both accuracy and computational time. Craig \& Sulem \cite{Ref17} proposed a perturbative approach for computing the DtN operator in the case of a flat bottom, assuming periodic lateral conditions. This strategy, based on a functional Taylor-series expansion of the DtN operator around the state of zero free-surface elevation, $\eta   (  {\bf x}  ,  {\kern 1pt} t  )    =    0$, has been further studied and advanced by many authors, see e.g. \cite{Ref45}, \cite{Ref46}, \cite{Ref47}, leading to useful numerical schemes for solving various problems. This approach, being of perturbative character, is very efficient when slightly-to-moderately deformed fluid domains are considered. Extensions to large variations of the bathymetry profile become complicated, and of increased computational cost \cite{Ref48}, \cite{Ref49}, because of keeping more modes in the corresponding expansions. It should be noted here, that a perturbative method may fail as the number of modes increases. See, e.g. \cite{Ref50}.

\subsection{A rapidly convergent ``modal'' series expansion of the wave potential }

Recently, an alternative formulation has been proposed, based on an exact (non-perturbative), rapidly convergent series expansion of the unknown wave potential $\Phi   (  \xh  ,  {\kern 1pt} z  ,  {\kern 1pt} t  )$ of the form

\begin{equation} \label{GrindEQ__3_}
\Phi   (  \xh  ,  {\kern 1pt} z  ,  {\kern 1pt} t  )      =\sum _{n    =    -    2}^{\infty }\varphi _{n} (  \xh  ,  t  )    Z_{n} \left({\kern 1pt} z  {\kern 1pt} ;  {\kern 1pt} \eta   (  \xh  ,  t  )  ,  h  (  x  )\right) .
\end{equation}
As proved in detail in AP17, the vertical basis system $\left\{Z_{-2} ,Z_{-1} ,\{   Z_{n}   \} _{n\ge 0} \right\}$, is composed of:
 \begin{enumerate}[(i)]
 	\item The two specific functions
 	\begin{subequations}\label{eq:Zadd}
 		\begin{align}
 		Z_{  -  2} (  z  ;  {\kern 1pt} \eta   ,  {\kern 1pt} h  )    &=    \frac{\mu _{  0}   h_{  0} +1}{2  {\kern 1pt} h_{  0} }   {\kern 1pt} \frac{(  z    +    h  )}{H} ^{{\kern 1pt} 2}       -      \frac{\mu _{  0}   h_{  0} +1}{2  h_{  0} }   {\kern 1pt} H      +      1,\\
 		Z_{  -  1} (  z  {\kern 1pt} ;  \eta   ,  h  )    &=    \frac{\mu _{  0}   h_{  0} -1}{2  h_{  0} }   {\kern 1pt} \frac{(  z    +    h  )}{H} ^{{\kern 1pt} 2}     +    {\kern 1pt} \frac{1}{h_{  0} }   {\kern 1pt} (  z    +    h  )    {\kern 1pt} -    {\kern 1pt} \frac{\mu _{  0}   h_{  0} +1}{2  h_{  0} }   {\kern 1pt} H    {\kern 1pt} +    {\kern 1pt} 1         
 		\end{align}
 	\end{subequations}
 	where $H    =    H(  \xh  ,  t  )    =    \eta   (  \xh  ,  t  )    +    h  ( \xh  )$ is the local depth of the fluid up to the instantaneous free surface, and $\mu _{  0}   ,  {\kern 1pt} h_{  0} $ are two auxiliary constants, which will be discussed in the sequel. These functions serve the purpose to free the expansion \eqref{GrindEQ__3_} of the boundary constraints imposed by the remaining part of the expansion (see Eqs. \eqref{ZnBCs}, below), resulting also in a significant acceleration of the convergence.
 	
 	\item    The set of local eigenfunctions $\{   Z_{n}     =    Z_{n}   (  {\bf x}  ,  {\kern 1pt} t  )  \} _{n    \ge     0} $ of a regular Sturm-Liouville problem, defined in the vertical interval $\left[-h  (  \xh  )  ,  {\kern 1pt} \eta   (  \xh  ,  t  )\right]$. In principle, any such problem can provide a theoretically acceptable system of local eigenfunctions. For physical reasons (see comments below), in constructing the HCMT, the choice has been made of the vertical Sturm-Liouville problem corresponding to linear water waves, with boundary conditions
 	\refstepcounter{equation}
 	$$
 		\frac{\partial {\kern 1pt} Z_{  n} }{\partial {\kern 1pt} {\kern 1pt} z}       -      \mu _{  0}   Z_{  n}       =      0  ,\,\,         {\rm at}\,\,        z    =    \eta   (  {\bf x}  ,  {\kern 1pt} t  )\label{ZnBCa}  ,\qquad
 	\frac{\partial {\kern 1pt} Z_{  n} }{\partial {\kern 1pt} {\kern 1pt} z}       =      0  ,\,\,          {\rm at}\,\,        z    =    -h  (  {\bf x}  ).\eqno{(\theequation{\mathit{a},\mathit{b}})}\label{ZnBCs}
 	$$  
\end{enumerate}
Eigenfunctions $\{   Z_{n}   \} _{n    \ge     0} $, normalized to take the value 1 at $z    =$ $\eta   (  x  ,  t  )$, are given by the equations:
$$
	Z_{0}     =    \frac{\cosh (  k_{0} (z    +    h  )}{\cosh (  k_{0}   H  )} ,\qquad
 Z_{n}     =    \frac{\cos (  k_{n} (z    +    h  )}{\cos (  k_{n}   H  )} ,   n    \ge     1,\eqno{(\theequation{\mathit{c},\mathit{d}})}\label{Zntrig}
$$  
where $k_{n}     =    k_{n}   (  x  ,  {\kern 1pt} t  )$, $n    \ge     0$ are the roots of the following transcendental equations:
\begin{subequations}\label{eq:kns}
	\begin{align}
	k_{  0} {\kern 1pt} H  {\kern 1pt} \tanh   (  k_{  0} {\kern 1pt} H  )    &=    \mu   (  \xh  ,  t  )\label{eq:kn0}\\
	k_{  n} {\kern 1pt} H  {\kern 1pt} \tan   (  k_{  n} {\kern 1pt} H  )    &=    -    \mu   (  \xh  ,  t  ),\,\,         \text{for}\,\,    n    \ge     1\label{eq:knn}
	\end{align}
\end{subequations}
 and $\mu   (  x  ,  t  )    =    \mu _{  0}   H(  x  ,  t  )$. From the theory of Sturm-Liouville problems, it is known that the system $\{   Z_{n}   \} _{n    \ge     0} $ is an $L^{  2} -$basis in each vertical interval $\left(-h  ,  {\kern 1pt} \eta \right)$ \cite{Ref51}. The addition of the two functions $Z_{  -  2}   ,  {\kern 1pt} Z_{  -  1} $ makes the extended system $\left\{Z_{-2} ,Z_{-1} ,\{   Z_{n}   \} _{n\ge 0} \right\}$ a basis in the Sobolev space $H^{  2} \left(-h  ,  {\kern 1pt} \eta \right)$, ensuring quick, point-wise convergence of the series itself to $\Phi   (  \xh  ,  {\kern 1pt} z  ,  {\kern 1pt} t  )$, and of the term-wise differentiated series to the corresponding derivatives of $\Phi   (  \xh  ,  {\kern 1pt} z  ,  {\kern 1pt} t  )$; see AP17 for a detailed proof. The additional basis functions, $Z_{  -  2}   ,  {\kern 1pt} Z_{  -  1} $, have been adopted by other researchers as well, who studied their role in the convergence of numerical schemes for solving the Helmholtz equation \cite{Ref52}, and ascertained their strong positive effects in solving problems in acoustical waveguides of irregular shape \cite{Ref53}, \cite{Ref54}. In our paper AP17, we have proved that, if $\eta $, $h$ and $\Phi $ are sufficiently smooth functions, then the ``modal'' amplitudes $\varphi _{n} (  \xh  ,  t  )$, and their derivatives $\nabla _{\xh} \varphi _{n} (  \xh  ,  t  )$, $\partial _{t} {\kern 1pt} \varphi _{n} (  \xh  ,  t  )$ decay as fast as $O  (  n^{-4}   )$, while the decay of the first few modes (-2, -1, 0, 1, 2, 3) is even faster, namely exponential. This suggests that only a few modes are enough for accurate computations, as also confirmed by numerical experiments.
 
 The auxiliary constant $h_{  0} $, appearing in Eqs. \eqref{eq:Zadd}, is introduced only for dimensional purposes, and its value is taken to be a characteristic depth of the studied configuration, e.g. the depth at the incident region, or the mean depth. As regards the auxiliary constant $\mu _{  0} $, the following comments are in order. The essential role of this constant is to formulate the first boundary condition in \eqref{ZnBCa} of the Sturm-Liouville problem defining the eigenfunctions $\{   Z_{n}   \} _{n    \ge     0} $. All theoretical statements made above remain valid for any value $\mu _{  0}     >    0$. Since, however, for a specific choice of $\mu _{  0} $, the eigenfunctions $\{   Z_{n}   \} _{n    \ge     0} $ become the physical modes of linear waves for angular frequency $\omega _{  0}^{}     =    \sqrt{{\kern 1pt} g{\kern 1pt} \mu _{  0} } $ (at the local depth), it is preferable to select $\mu _{  0} $ in relation with a characteristic frequency of the problem, e.g. the frequency of the incident wave, or the central frequency of a wave packet or wave spectrum. In this way, the series expansion \eqref{GrindEQ__3_} encapsulates the physics of a ``nearby'' linear wave problem, even before using the nonlinear Hamiltonian equations (see Eqs. \eqref{GrindEQ__7_}), by means of which the solution will be determined taking fully into account all nonlinear features of the problem. For reasons explained above, this is not an approximation; it is, instead, an adaptation of the vertical expansion, making it converge even faster, since it takes \textit{a priori} into account a part of the physical structure of the problem.

 \textbf{Comment on terminology}. The two first terms of Eq. \eqref{GrindEQ__3_}, $\varphi _{-2} {\kern 1pt} Z_{-2} $ and $\varphi _{-1} {\kern 1pt} Z_{-1} $, are called, respectively, \textit{free-surface mode} and \textit{sloping bottom mode}, since they ensure the ability of the series to correctly satisfy the free-surface and bottom conditions. Also, they are referred to collectively as \textit{boundary modes}. The term $\varphi _{  0} {\kern 1pt} Z_{  0} $ is referred to as the \textit{propagating mode}, and the remaining terms, $\varphi _{  n} {\kern 1pt} Z_{  n} $ for $n    \ge     1$, are called \textit{evanescent modes}, by analogy with the linear wave theory. It should be stressed, however, that the individual terms of the series \eqref{GrindEQ__3_} have not the usual meaning of modes (as in the linear wave propagation), since all terms together solve the nonlinear hydrodynamic problem. This is why we have used the word mode in quotes up to now, a convention which will be abandoned in the sequel, after these explanations.

 \subsection{The HCMT}

Introducing the series expansion \eqref{GrindEQ__3_} of the wave potential into Luke's variational principle \cite{Ref55}, and performing the variations with respect to the unknown fields $\eta   (  \xh  ,  {\kern 1pt} t  )$ and $\varphi _{  n} (  \xh  ,  {\kern 1pt} t  )$, ${\kern 1pt} n    \ge     -2$, we obtain an infinite set of Euler-Lagrange equations. Elaborating further on this set of equations, we find the following two evolution equations with respect to $\eta   (  \xh  ,  {\kern 1pt} t  )$ and $\psi (  \xh  ,  {\kern 1pt} t  )    =$ $\sum _{  n    \ge     -  2  }\varphi _{  n} (  \xh  ,  {\kern 1pt} t  ) $,
\begin{subequations}\label{GrindEQ__7_}
	\begin{align}
	\partial _{{\kern 1pt} t} \eta    & =    -    (  \nabla _{\xh}^{} \eta   )\cdot (  \nabla _{\xh}^{} \psi   )    +    (  {\kern 1pt} |\nabla _{\xh}^{} \eta {\kern 1pt}   |^{  2}   +  1  )    \left(  h_{  0}^{  -    1}   {\kern 1pt} \varphi _{-2}     +    \mu _{  {\kern 1pt} 0}   \psi   \right),\\
	\partial _{{\kern 1pt} t} \psi    & =    -    g{\kern 1pt} \eta     -    \frac{1}{2} (  \nabla _{\xh}^{} \psi   )^{  2} +    \frac{1}{2} ({\kern 1pt}   |\nabla _{\xh}^{} \eta {\kern 1pt}   |^{  2}   +  1  )    \left(  h_{  0}^{  -   1}  {\kern 1pt} \varphi _{-2}  + \mu _{ 0}  \psi  \right)^{ 2},
	\end{align}
\end{subequations}
coupled, through the field $\varphi _{-2}     =    \varphi _{-2} (  \xh  ,  {\kern 1pt} t  )$, with an infinite set of time-independent equations (with coefficients parametrically dependent on time) with respect to $\varphi _{  n} (  \xh  ,  {\kern 1pt} t  )  ,  {\kern 1pt} n    \ge     -2$, given by
\begin{subequations}\label{GrindEQ__8_}
	\begin{align}
	\sum _{n    =    -2}^{\infty }\left(A_{m  n}^{} \nabla _{\xh}^{2}     +    B_{m  n}^{} \cdot \nabla _{\xh}^{}     +    C_{{\kern 1pt} m{\kern 1pt} n}^{} \right)  \varphi _{n}     &=      0  ,\quad                m\ge -2,\\
	\sum _{n    =    -2}^{\infty }{\kern 1pt} \varphi _{n}       & =      \psi.
	\end{align}
\end{subequations}
The $(  \xh  ,  t  )-$ dependent matrix coefficients $A_{m  n}^{} =A_{m n}( \eta, h )$, $\mathbf{B}_{m  n}^{} =\left(B_{m  n}^{1} (  \eta   ,  h  ),   B_{m  n}^{2} (  \eta   ,  h  )\right)$ and $C_{m  n}^{} =C_{m  n}^{} (  \eta   ,  h  )$ are defined in terms of $\eta   (  x  ,  t  )$ and $h  (  \xh  )$ through the vertical functions $Z_{  n}     =Z_{  n} \left( z   ;   \eta   (  \xh  ,  t  )  ,  h  (  \xh  )\right)  ,\,\,      n    \ge     -2$, by the equations
\begin{subequations}\label{ABCs}
	\begin{align}
	A_{{\kern 1pt} m{\kern 1pt} n}^{} &=\int _{-h}^{  \eta }Z_{n}   Z_{m}   d{\kern 1pt} z,\\
	\mathbf{B}_{m{\kern 1pt} n}^{} &=    2\int _{-h}^{  \eta }(\nabla _{\xh}^{} Z_{n}   )  Z_{m}   d{\kern 1pt} z     +    (\nabla _{\xh}^{} h  )  \left[  Z_{m}   Z_{n}   \right]_{  z  =  -  h},\\
	C_{{\kern 1pt} m{\kern 1pt} n}^{}       &=\int _{-h}^{  \eta }(  \nabla _{\xh}^{2} {\kern 1pt} Z_{n}     +    \partial _{{\kern 1pt} z}^{{\kern 1pt} 2} {\kern 1pt} Z_{n}   )  {\kern 1pt} Z_{m}   {\kern 1pt} d{\kern 1pt} z     -    N_{h}   \cdot   {\kern 1pt} \left[  (\nabla _{\xh}^{} Z_{{\kern 1pt} n}   ,  {\kern 1pt} \partial _{{\kern 1pt} z} Z_{{\kern 1pt} n}   )  {\kern 1pt} Z_{m}   \right]_{  z  =  -  h}.
	\end{align}
\end{subequations}
Equations \eqref{GrindEQ__8_} describe the kinematics of the fluid (substrate kinematics) at each time, for given bathymetry $h  (  x  )$, free-surface elevation $\eta   (  x  ,  t  )$ and free-surface potential $\psi   (  x  ,  t  )$. They form an elliptic problem which should be supplemented by appropriate lateral boundary conditions, dependent on the specific problem considered. A detailed description of the lateral boundary conditions for various specific problems, both in 2D and in 3D configurations, can be found in the Supplementary Material (Appendix C) of the paper \cite{Ref56}, and in \cite{Ref57}, available online through the link \url{https://arxiv.org/abs/1710.10847} .

In the above formulation, Eqs. \eqref{GrindEQ__7_} bear in mind the classical Hamiltonian formulation, Eqs. \eqref{GrindEQ__1_} -- \eqref{GrindEQ__2_}, with $\varphi _{-2} (  x  ,  {\kern 1pt} t  )$ being a nonlocal coefficient (different, yet) related with the DtN operator ${\rm {\mathcal G}}  {\kern 1pt} [  \eta   ,  {\kern 1pt} h  ]  {\kern 1pt} \psi $ through the equation
\begin{equation} \label{GrindEQ__10_}
{\rm {\mathcal G}}  {\kern 1pt} [  \eta   ,  {\kern 1pt} h  ]  {\kern 1pt} \psi       =      -    (  \nabla _{x} \eta   )  \cdot   {\kern 1pt} (  \nabla _{x} \psi   )    +    (  {\kern 1pt} |\nabla _{x} \eta {\kern 1pt}   |^{  2}   +  1  )    \left(  h_{  0}^{  -    1}   {\kern 1pt} \varphi _{-2}     {\kern 1pt} +{\kern 1pt}     \mu _{{\kern 1pt} 0}   \psi   \right).
\end{equation}
Eq. \eqref{GrindEQ__10_} is obtained by comparing Eqs. \eqref{GrindEQ__1_} and \eqref{GrindEQ__2_} with Eq. \eqref{GrindEQ__7_}; as has also been established independently in AP17. Further, as proved in the same paper, the approximation of $\varphi _{-2} (  \xh  ,  {\kern 1pt} t  )$, and thus of the DtN operator through Eq. \eqref{GrindEQ__10_}, resulting from solving the (truncated version of the) modal kinematical problem, Eq. \eqref{GrindEQ__8_}, turns out to be amazingly effective for any (smooth) bathymetry. This is due to the fact that the truncated problem \eqref{GrindEQ__8_} exhibits a miraculous superconvergence with respect to the number of modes $N_{{\rm tot}}^{} $ kept in the truncated expansion; the error diminishes at a rate proportional to $O  (  N_{{\rm tot}}^{  -  6.5}   )$, for any shape (however steep) of the free-surface and bottom boundaries. Accordingly, a small number of modes suffices to provide very accurate results for the modal amplitude $\varphi _{-2} (  \xh  ,  {\kern 1pt} t  )$, making Eqs. \eqref{GrindEQ__7_} essentially exact. This is also confirmed by the examples provided herein, in Section \ref{sec:6}, as well as by other simulations concerning demanding problems with strong nonlinearity and dispersion; see e.g. \cite{Ref1}, \cite{Ref16}, \cite{Ref57}.

\section{Implementation of the HCMS}
\label{sec:3}
\subsection{Scheme of numerical solution}
In order to solve the evolution Eqs. \eqref{GrindEQ__7_}, we need to calculate the nonlocal coefficient $\varphi _{-2} (  \xh  ,  t  )$ at each time step. That is, we have to solve the elliptic problem \eqref{GrindEQ__8_} for given $\eta   (  \xh  ,  t  )$ and $\psi   (  \xh  ,  t  )$ (known from the previous iteration, or from the initial conditions). For this purpose, system \eqref{GrindEQ__8_} is truncated at a finite number of modes $N_{{\rm tot}}     =    M    +    3$, where $M$ is the order of the last kept evanescent mode, and discretised by using central finite differences of fourth-order accuracy. For example, in the case of one horizontal dimension $(  \partial _{x_{2} }     \equiv     0  )$ and periodic lateral conditions, the discretised version of Eqs. \eqref{GrindEQ__8_} reads as follows:
	\begin{subequations}
			\begin{align}
			\sum _{n =-  2}^{M} \mathsmaller{\left(  -  \frac{A_{  m{\kern 1pt} n}^{i} }{12  \Delta x^{2} }     +    \frac{B_{  m{\kern 1pt} n}^{i} }{12  \Delta x}   \right)}\varphi _{n}^{i    -    2}     +     \mathsmaller{\left(  \frac{4A_{  m{\kern 1pt} n}^{i} }{3  \Delta x^{2} }     -    \frac{2B_{  m{\kern 1pt} n}^{i} }{3  \Delta x}   \right) } {\kern 1pt} \varphi _{n}^{i    -    1} &    - \mathsmaller{\left(  \frac{5A_{  m{\kern 1pt} n}^{i} }{2  \Delta x^{2} }     -    C_{  m{\kern 1pt} n}^{i} \right) }\varphi _{n}^{i}     \nonumber\\
			+    \mathsmaller{\left(  \frac{4A_{  m{\kern 1pt} n}^{i} }{3  \Delta x^{2} }     +    \frac{2B_{  m{\kern 1pt} n}^{i} }{3  \Delta x}   \right)}\varphi _{n}^{{\kern 1pt} i    +    1}     -\mathsmaller{\left(\frac{A_{  m{\kern 1pt} n}^{i} }{12  \Delta x^{2} }     +    \frac{B_{  m{\kern 1pt} n}^{i} }{12  \Delta x}   \right)}      {\kern 1pt} \varphi _{n}^{i    +    2}      & =      0,\quad   \left\{\begin{array}{l}
			\mathsmaller{i=1,...,N_{X}} ,\\
			\mathsmaller{m    =    -2,...,N_{{\rm tot}} -1}  ,
			\end{array}\right.\\
			\sum _{n    =    -  2}^{M}\varphi _{n}^{i}    &  =    \psi ^{i} ,\quad      i=1,...,N_{X}
			\end{align}
	\end{subequations}
where $\Delta {\kern 1pt} x$ is the mesh size, $N_{X} $ is the number of grid points, and the upper index $(  {\kern 1pt} \bullet   ^{i}   )$ denotes discrete local values; e.g. $\varphi _{n}^{{\kern 1pt} i  }     =    \varphi _{n}^{} (  x_{i}   ,  {\kern 1pt} t  )$ and $A_{  m{\kern 1pt} n}^{i}     =    A_{  m{\kern 1pt} n}^{} (  \eta   (  x_{i}   )  ,  {\kern 1pt} h  (  x_{i}   )  )$. The convergence and accuracy of the above scheme is investigated in AP17; see also \cite{Ref58}.

Having solved the above linear system, the local values of the free surface mode $\varphi _{-2}^{{\kern 1pt} i  }     =$ $\varphi _{-2}^{} (  x_{i}   ,  {\kern 1pt} t  )  ,$ $i=1,...,N_{X} $, are extracted and used in the evolution Eqs. (7a,b). Introducing the vector field $U    =    (  \eta   ,  {\kern 1pt} \psi   )^{  {\rm T}} $, system \eqref{GrindEQ__7_} is written as $\partial _{t} U    =    N(  t  ,  {\kern 1pt} U  )$ where$N(  t  ,  {\kern 1pt} U  )$ is the vector function defined by the right-hand sides of Eqs. (7a,b). Using now a temporal discretization $t^{  n} $, $n    =    1  ,  {\kern 1pt} ...  {\kern 1pt} ,N_{T}   ,$ and the notation $U^{n}     =$ $U(  x  ,  t^{  n}   )    =$ $(  \eta   (  x  ,  t^{  n}   )  ,  {\kern 1pt} \psi   (  x  ,  t^{  n}   )  )^{{\rm T}} $, system \eqref{GrindEQ__7_} is marched in time by a straightforward adaptation of the classical Runge-Kutta method, based on the Butcher tableau RK41 given in \cite[p. 91]{Ref59}. The whole scheme of numerical solution of the HCMS, Eqs. \eqref{GrindEQ__7_} and \eqref{GrindEQ__8_}, is presented in Algorithm \ref{algorithm1}.
\begin{center}
	\begin{algorithm}
		\caption{Numerical solution of the HCMS, Eqs. \eqref{GrindEQ__7_}, \eqref{GrindEQ__8_}}
		\label{algorithm1}
		\begin{algorithmic}
			\State
			\State{Given $U^{0}     =    (  \eta ^{0}   ,  \psi ^{0}   )$,\\
				\\
				Calculate $k_{n} (  \eta ^{0} ,h  )$, Eqs. \eqref{eq:kns}, and $A  (  \eta ^{0}   ,  h  )$, $B  (  \eta ^{0}   ,  h  )$, $C  (  \eta ^{0}   ,  h  )$, Eqs. \eqref{ABCs}. \\
				\\
				Solve CMS, Eqs. \eqref{GrindEQ__8_}, with $\psi     =    \psi ^{0} $, to find $\varphi _{-2}^{} (  x  ,  t )$}
			\State
			\For{$m=1\rightarrow N_{T}$}
			\State
			\For{$l=1\rightarrow 4$}
			\State
			\State{$\bar{U}^{l} _{}     \equiv     (  \bar{\eta }^{l} _{}   ,  \bar{\psi }^{l} _{}   )^{{\rm T}}     =    U^{m}     +    \delta t  \sum _{j    =    1}^{l  -  1}a_{{\kern 1pt} l{\kern 1pt} j}   K_{{\kern 1pt} j}^{m}$ }
			\State
			\State{Calculate $k_{n} (  \bar{\eta }^{l} _{} ,h  )$, Eqs. \eqref{eq:kns}, and $A(  \bar{\eta }_{}^{l}   ,  h  )$, $B  (  \bar{\eta }_{}^{l}   ,  h  )$, $C(  \bar{\eta }_{}^{l}   ,  h  )$, Eqs. \eqref{ABCs}}
			\State
			\State{and Solve CMS, Eqs. \eqref{GrindEQ__8_}, with $\psi     =    \bar{\psi }^{{\kern 1pt} l} $, to find $\varphi _{-2}^{} (  x  ,  t  )$}
			\State
			\State{	$K_{l}^{m}     =    N(  t^{m}     +    c_{l}   \delta t  ,  {\kern 1pt} \bar{U}^{l}   )$}
			\State
			\EndFor
			\State
			\State{$U^{m+1} \, \, \, \equiv \, \, \, (\, \eta ^{\, m\, +\, 1} ,\, {\kern 1pt} \psi ^{\, m\, +\, 1} \, )\, \, \, =\, \, \, U^{m} \, \, +\, \, \delta t\, {\kern 1pt} \sum _{j\, \, =\, \, 1}^{4}b_{{\kern 1pt} j} \, K_{{\kern 1pt} j}^{\, m}$}
			\EndFor
		\end{algorithmic}
	\end{algorithm}
\end{center}

 It should be noted that Algorithm \ref{algorithm1} requires four evaluations of the matrix coefficients $A    =    \left(A_{{\kern 1pt} m{\kern 1pt} n}^{} {\kern 1pt} \right)$, $B    =    \left({\kern 1pt} \left(B_{{\kern 1pt} m{\kern 1pt} n}^{  {\kern 1pt} (1)} \right)  ,  {\kern 1pt} \left(B_{{\kern 1pt} m{\kern 1pt} n}^{  {\kern 1pt} (2)} \right){\kern 1pt} \right)$ and $C    =    \left(C_{{\kern 1pt} m{\kern 1pt} n}^{} \right)$ at each point of the spatial grid, for the implementation of one time step. Thus, a huge number of evaluations of these coefficients is required, which may count up to $10^{  11} $ times for a long-time simulation. Accordingly, the fast and accurate calculation of all elements of these coefficients is of fundamental importance for the efficient implementation of the HCMS. Modifications of the above numerical scheme for solving other variants of water-wave problems, e.g. problems with different boundary conditions (vertical impermeable walls or generating and absorbing layers) are possible, and have been also developed. More details can be found in \cite[Chapter 7]{Ref58} and \cite{Ref60}.
 
 \subsection{Analytic calculation of the $(\xh , t )-$varying matrix coefficients $A$,${\bf B}$ and ${\kern 1pt} C$}
 
 The best approach to ensure fast and accurate evaluations of all elements of the matrix coefficients $A$, ${\bf B}$, $C$, is to derive closed-form analytic expressions of them. This requires extensive analytic manipulations, which have been performed and tested against numerical integration using Eqs. \eqref{ABCs}. A systematic account of these calculations can be found in \cite{Ref58}. The main findings are: (i) all coefficients can be expressed in closed form, in terms of $h_{  0} $, $\mu _{  0}   ,  {\kern 1pt} \eta $, $h$, $k_{n} \left(\eta   ,  {\kern 1pt} h\right)$, and (ii) using these closed forms, accurate evaluations of the coefficients becomes about 150 times faster than by using numerical integration, at the same level of accuracy. As an example, we give here the closed form expressions of the coefficients $A_{  m  n} $ and $\mathbf{B}_{  mn} $ for the case $m  {\kern 1pt} ,  {\kern 1pt} n    \ge     1$:
\begin{subequations}\label{eq:ABCsAn}
	\begin{align}
	A_{  m  n}       &=      \left\{\begin{array}{l} {    \frac{H}{2}     +  \frac{H\mu _{0} ^{2}     -    \mu _{0} }{2k_{n} ^{2} }   ,                  m    =    n    \ge     1  ,} \\ {                            0  ,\quad m    \ne     n  ,} \end{array}\right.\\
	\bf{B}_{  m{\kern 1pt} n}^{}    & =      2{\kern 1pt}   \left(\mathbf{F}_{n}^{  (  1  )}   I_{m  n}^{  (  1  )}     +    \mathbf{F}_{n}^{  (  2  )}   I_{m  n}^{  \eqref{GrindEQ__2_}}     +    \mathbf{F}_{n}^{  (  3  )}   I_{m  n}^{  (  3  )} \right)    +    (  \nabla _{\xh}^{} {\kern 1pt} h  )  \left[Z_{  n}   Z_{  m} \right]_{  z    =    -  h},
	\end{align}
\end{subequations}
with
\[\mathbf{F}_{n}^{  (  1  )}     =    -\mu _{0}   \left(  1    +    H  {\kern 1pt} \frac{\partial _{H} k_{n} }{k_{n} }   \right)  \nabla _{\xh}^{} H,\quad  I_{m  n}^{  (  1  )}     =    A_{  m  n} ,  \]
\[\mathbf{F}_{n}^{  (  2  )}       =      -    k_{n}   \nabla _{\xh}^{} h,\quad  I_{m  n}^{  (  2  )}     =      \left\{\begin{array}{l} {                                                      \frac{\mu _{0}^{2} }{2  k_{m}^{3} }     ,                                                                            m    =    n    \ge     1  ,} \\ {  -    \frac{  k_{n}^{-  1}   \mu _{0}^{2}     +    k_{n}     -    k_{n} \left[  Z_{n}^{}   Z_{m}   \right]_{  -h}^{} }{k_{n}^{2}     -    k_{m}^{2} }   ,                    m\ne n  ,} \end{array}\right. \]

\[\mathbf{F}_{n}^{(  3  )}       =      -    \partial _{H} k_{n}   \nabla _{\xh}^{} H,\quad  I_{n}^{  (  3  )}     =      \left\{\begin{array}{l} {    \frac{-  \mu _{0}     +    H  (  \mu _{0}^{2}     -    k_{m}^{2}   )}{4  k_{m}^{3} }     ,                                              m    =    n    \ge     1  ,} \\ {    -    \frac{  k_{n}^{-  1}   \mu _{0}   (  H  \mu _{0}^{}     -1)    +    k_{n}   H}{k_{n}^{2}     -    k_{m}^{2} }     ,                    m\ne n  ,} \end{array}\right. \]
where the derivative of $k_{ n} $ with respect to $H = \eta+h$ is obtained by applying the implicit function theorem to the defining relation, Eq. \eqref{eq:knn}, and reads as follows:

\begin{equation} \label{GrindEQ__13_}
\partial _{H} {\kern 1pt} k_{ n}= \frac{k_{{\kern 1pt} n} \, (\, k_{{\kern 1pt} n}^{ 2} + \mu _{{\kern 1pt} 0}^{ 2}  ) }{ \mu _{{\kern 1pt} 0}^{} - H ( k_{{\kern 1pt} n}^{ 2}  + \mu _{{\kern 1pt} 0}^{ 2}  )} ,\quad         n \ge  1.
\end{equation}

 \subsection{Semi-analytic calculation of the roots $k_{  n} (  {\bf x}  ,  {\kern 1pt} t  )$ of the local dispersion relation }

 Having established explicit formulae for all coefficients $A_{{\kern 1pt} m{\kern 1pt} n}^{} $, $\mathbf{B}_{{\kern 1pt} m{\kern 1pt} n}^{} $ and ${\kern 1pt} C_{{\kern 1pt} m{\kern 1pt} n}^{} $ in terms of $h_{  0} $, $\mu _{  0}   ,  {\kern 1pt} \eta   ,$$h$ and $k_{n} \left(\eta   ,  {\kern 1pt} h\right)    =$ $k_{  n} (  {\bf x}  ,  {\kern 1pt} t  )$, with $h_{  0} $,$\mu _{  0}   ,  {\kern 1pt} \eta   ,  {\kern 1pt} h$ already known ($\eta $ is known from the previous time step), the question of their fast, accurate and robust evaluation is reduced to the corresponding question for the roots $k_{  n} (  {\bf x}  ,  {\kern 1pt} t  )$ of the local dispersion relations, Eqs. \eqref{eq:kns}. Solving these equations by means of the Newton-Raphson method, or any other iterative procedure, is in principle a trivial task. However, since this evaluation is to be performed for $10^{  8} -10^{  11} $ times, it is imperative the convergence of the solution scheme to be a priori ensured, and the number of iterations to remain small. For that reasons, it is important to establish highly accurate initial values for all $k_{  n} $'s or, even better, to derive satisfactory closed-form approximations.
 
 \textbf{For the case of} $k_{  0} $, explicit approximations, accurate up to the third or fourth decimal digit, for all values of the parameter $\mu     =    \mu   (  {\bf x}  ,  t  )    =$ $\mu _{  0} {\kern 1pt} H(  x  ,  t  )$ are available in the literature; see e.g. \cite{Ref30}, \cite{Ref31}, \cite{Ref32}, \cite{Ref33}, \cite{Ref61}, \cite{Ref62}. In this work, the approximation of \cite{Ref33} is used, which is based on Newton-Raphson iterations of the form
\begin{equation} \label{GrindEQ__14_}
{}^{j+1} \kappa _{  0}^{}   (  \mu   )      =      \frac{{}^{j} \kappa _{  0}^{2}   (  \mu   )    +    \mu \cosh ^{2} \left({}^{j} \kappa _{  0}^{}   (  \mu   )\right)}{{}^{j} \kappa _{  0}^{}   (  \mu   )    +    0.5  {\kern 1pt} {\kern 1pt} \sinh \left(2  {}^{j} \kappa _{  0}^{}   (  \mu   )\right)} ,
\end{equation}
with initial guess
\begin{equation} \label{GrindEQ__15_}
^{0} \kappa _{  0}^{}   (  \mu   )      =      \frac{\mu     +    \mu ^{  1.986} \exp   (-  {\kern 1pt} 1.863    +    1.198  {\kern 1pt} \mu ^{  1.366} )}{\sqrt{  \tanh \mu } } .
\end{equation}
The initiation formula \eqref{GrindEQ__15_} has maximum relative error of approximately $10^{  -    4} $, and the methodology leads to relative errors of less than $10^{  -    15} $ for any value of $\mu $, within 2 iterations.

\textbf{For the case of} $k_{  n}   ,  {\kern 1pt} n    \ge     1$, a single, simple and sufficiently accurate approximation, valid for all values of $\mu     =    \mu _{  0}   H(  \xh  ,  t  )$, is apparently not available. In addition, the application of the Newton-Raphson method to the solution of Eq. \eqref{eq:knn} is more involved, especially for large values of $\mu $ and small values of $n$, where, the steep gradient of tan function may lead to a large value of iterations, or to non-convergence or, even worse, to convergence to an erroneous solution. Sections \ref{sec:4} and \ref{sec:5} of this paper are devoted to the derivation, analysis and application of new, highly accurate approximations for the roots of Eq. \eqref{eq:knn}, that optimise the computation of $k_{n} $, $n    \ge     1$, for all values of $\mu $ and $n$, and, when used as initialization of an iterative solver, lead to machine precision tolerance $10^{  -    15} $ with no more than 3 iterations.

 \section{Root Approximation Strategy}
\label{sec:4}
 Eq. \eqref{eq:knn} can be written in the form
\begin{equation} \label{GrindEQ__16_}
g  (\kappa   {\kern 1pt} ;  {\kern 1pt} \mu   )    \equiv     \kappa \tan   (\kappa )    +    \mu       =      0,
\end{equation}
 where $\kappa _{n}     =    k_{n} (  x  ,  t  )  H  (  x  ,  t  )$ and $\mu     =    \mu   (  {\bf x}  ,  t  )    =    \mu _{  0}   H  (  {\bf x}  ,  t  )$. The first three roots of Eq. \eqref{GrindEQ__16_} are depicted in Fig. 1, for three different values of the parameter $\mu     =    0.1  ,  {\kern 1pt} 1.0  ,  {\kern 1pt} 5.0$. $\mathrm{T}$he difficulties in obtaining efficient root-finding formulae for Eq. \eqref{GrindEQ__16_} are well demonstrated in this figure. \textbf{First}, the roots $\kappa _{n} $ are dependent on the parameter $\mu $ which, in the context of HCMT, is a variable quantity, $\mu     =    \mu _{  0}   H(  x  ,  t  )$ spanning a wide range of values, since the method is applied to all depths and wave amplitudes. Recall also that $\mu _{  0} $ is an auxiliary numerical parameter, arbitrarily chosen, although some general principles for its appropriate (but nonunique) selection have been discussed in Section 2.2. \textbf{Second}, the few first roots, which are the most important for applications, are associated to large values and steep gradients of the $\tan $ function, especially for large values of $\mu $. Accordingly, the obvious initial choice ${}^{0} \kappa _{  n}     =    n  \pi $, which is the asymptotic limit of $\kappa _{  n} $ for $n\to \infty $, is not appropriate in these cases, and may lead to a large number of Newton-Raphson iterations, or to nonconvergence or, even, to erroneous solutions.
 
 Beginning with the observation that for any $n\in \mathbb{N}$ the roots of Eq. \eqref{GrindEQ__16_} satisfy 
 \begin{equation}
 \kappa _{n} \in \left(\frac{(2n-1)\pi }{2} ,  \frac{(2n+1)\pi }{2} \right)\quad      \text{and}\quad      \lim \kappa _{n} \to n\pi ,   \quad\text{for all}\quad \mu     >    0, 
 \end{equation}
and following Newman \cite{Ref28}, we introduce the quantity
\begin{equation} \label{GrindEQ__18_}
\varepsilon _{  n} (\mu )      =      n\pi     -    \kappa _{  n} (\mu ).
\end{equation}
Clearly, $\varepsilon _{  n}   (  \mu   )$ satisfy the following relations:
\begin{equation} \label{GrindEQ__19_}
\varepsilon _{  n}   (  \mu   )    {\kern 1pt} \in     {\kern 1pt} \left(0  ,  {\kern 1pt} \pi /2\right),\quad        \mathop{\lim }\limits_{n    \to     \infty }   \varepsilon _{  n}   (  \mu   )    \to     0,\quad        \mathop{\lim }\limits_{\mu     \to     \infty }   \varepsilon _{  n}   (  \mu   )    \to     0.
\end{equation}
The values $\varepsilon _{n} (\mu )$ are monotonically decreasing with respect to $n$ for each $\mu  > 0$, and monotonically increasing with respect to $\mu $ for each $n {\kern 1pt} \in  {\kern 1pt} {\rm {\rm N}}$. The relation between $\kappa _{n} (\mu )$ and $\varepsilon _{n} (\mu )$ is depicted in Fig. 2.

Using Eq. \eqref{GrindEQ__18_}, Eq. \eqref{GrindEQ__16_} is written as $(n\pi -\varepsilon )\tan (n\pi -\varepsilon )=-\mu $, which, upon expanding $\tan (n\pi -\varepsilon )$, becomes
\begin{equation} \label{GrindEQ__20_}
(n\pi -\varepsilon )\tan (\varepsilon )=\mu .
\end{equation}

\begin{figure*}
	\begin{center}
		\includegraphics*[width=1.0\textwidth]{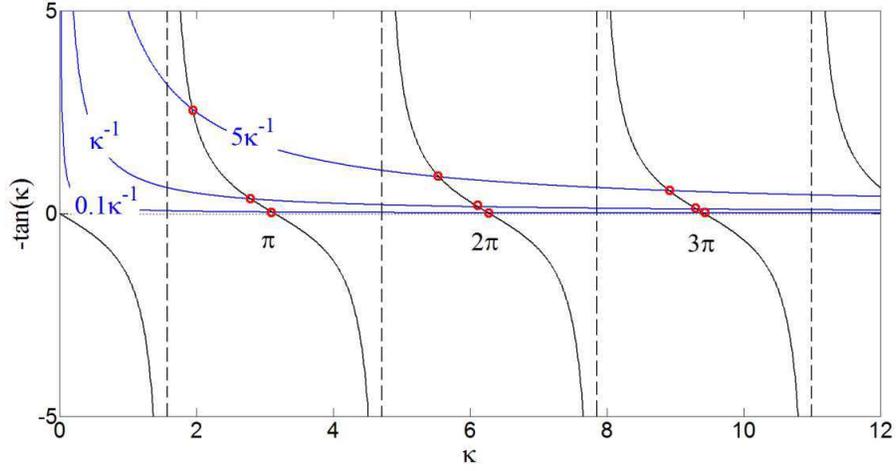}
		\caption{The first three roots of the transcendental Eq. \eqref{eq:knn}, for three different values of the parameter $\mu $ ($\mu     =    0.1  ,  {\kern 1pt} 1.0  ,  {\kern 1pt} 5.0$)}
		\label{fig:1}       
	\end{center}
\end{figure*}
By rewriting Eq. \eqref{GrindEQ__20_} in the form $\tan (\varepsilon )=\mu /\kappa $, a Picard iteration is readily defined as follows
\begin{equation}\label{GrindEQ__21_}
{}^{j+1} \varepsilon _{n} ={\rm Arctan}\left(\frac{\mu }{{}^{j} \kappa _{n} } \right),\,\,\text{for}\,\, j=0,\, 1,\, 2,\, ...\,\,\text{and an initial estimation}\,\,  ^{0} \varepsilon _{n},
\end{equation}
where $^{j} \kappa _{n} =n\pi -{}^{j} \varepsilon _{n} $. For later use we also write down the formula
\begin{equation} \label{GrindEQ__22_}
\cot \left({}^{j+1} \varepsilon _{n} \right) = {}^{j} \kappa _{n} /\mu ,
\end{equation}
which is, of course, equivalent with Eq. \eqref{GrindEQ__21_}. On the other hand, simple algebraic manipulations show that the function $g  (  \kappa {\kern 1pt}   ;  {\kern 1pt} \mu )$, Eq. \eqref{GrindEQ__16_}, can be written in the form $g  (  \kappa =n\pi -\varepsilon {\kern 1pt}   ;  {\kern 1pt} \mu )    {\kern 1pt} =$ ${\kern 1pt} -f(\varepsilon   {\kern 1pt} ;\mu )/\tan   (\varepsilon )$, where
\begin{equation} \label{GrindEQ__23_}
f(\varepsilon   {\kern 1pt} ;\mu )      =      n\pi -\varepsilon -\mu   {\kern 1pt} \cot   (\varepsilon ).
\end{equation}
\begin{figure*}
	\begin{center}
		\includegraphics*[width=1.0\textwidth]{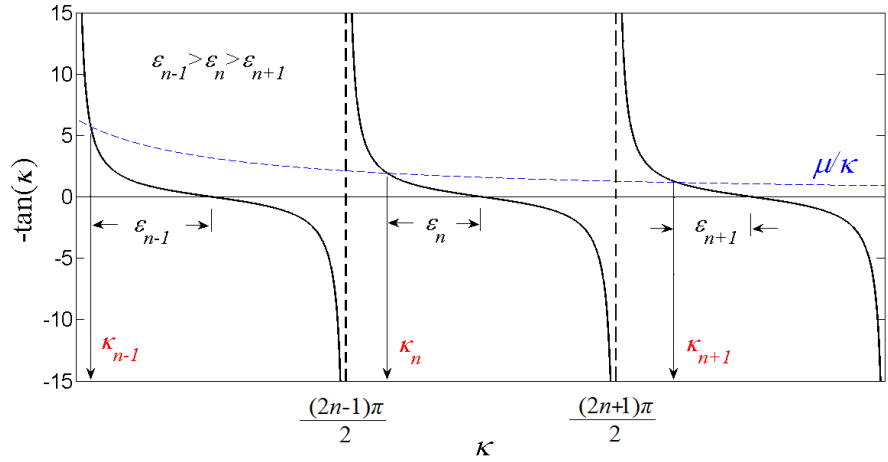}
		\caption{Definition of the sequence $\varepsilon _{n} (\mu )$}
		\label{fig:2}       
	\end{center}
\end{figure*}
Since $\tan   (\varepsilon _{  n} )    \ne     0  ,  {\kern 1pt} \pm \infty $, the solutions $\varepsilon _{  n} $ of Eq. \eqref{GrindEQ__21_} can also be found as the roots of the function $f(\varepsilon   {\kern 1pt} ;\mu )$. A Newton-Raphson procedure for the determination of $\varepsilon _{n} $ such that $f(\varepsilon _{n}   ;\mu )=0$, is given by the formula
\begin{equation} \label{GrindEQ__24_}
{}^{j+1} \varepsilon _{n}       =      ^{j} \varepsilon _{n}     -    \frac{f  (  ^{j} \varepsilon _{n}   ;  \mu   )}{\partial _{\varepsilon } f  (  ^{j} \varepsilon _{n}   ;  \mu   )} ,\,\,\text{for}\,\, j=0,  1,  2,  ...\,\,\text{and an initial estimation}\,\,     ^{0} \varepsilon _{n}, 
\end{equation}
which specializes in the form
\begin{equation} \label{GrindEQ__25_}
{}^{j+1} \varepsilon _{n}       =      ^{j} \varepsilon _{n}     -    \frac{n\pi     -    ^{j} \varepsilon _{n}     -    \mu   {\kern 1pt} \cot   (  {}^{j} \varepsilon _{n} )}{\mu     -    1    +    \mu   co{\kern 1pt} t^{2}   (  {}^{j} \varepsilon _{n} )} ,
\end{equation}
upon substituting $f    =    f(\varepsilon   {\kern 1pt} ;\mu )$ from Eq. \eqref{GrindEQ__23_} to Eq. \eqref{GrindEQ__24_}. Further, a third-order formula can be obtained by employing the Householder method \cite[Section 4.4]{Ref63}, having the form
\begin{equation} \label{GrindEQ__26_}
{}^{j+1} \varepsilon _{n}       =      ^{j} \varepsilon _{n}     -    \frac{f  (  ^{j} \varepsilon _{n}   ;  \mu )}{\partial _{\varepsilon } f  (  ^{j} \varepsilon _{n}   ;  \mu )}     \left[  1    +    \frac{f  (  ^{j} \varepsilon _{n}   ;  \mu )    \partial _{\varepsilon \varepsilon } f  (  ^{j} \varepsilon _{n}   ;  \mu )}{2    \left(\partial _{\varepsilon } f  (  ^{j} \varepsilon _{n}   ;  \mu )\right)^{2} }   \right].
\end{equation}
The above iteration equation takes the specific form
\begin{multline} \label{GrindEQ__27_}
{}^{j+1} \varepsilon _{n}       =        ^{j} \varepsilon _{n}     -    \frac{n\pi     -    ^{j} \varepsilon _{n}     -    \mu   {\kern 1pt} \cot   (  {}^{j} \varepsilon _{n} )}{\mu     -    1    +    \mu   \cot^{2}   (  {}^{j} \varepsilon _{n} )}\times\\     \times\left[  1    -    2  \mu     \frac{\left(n\pi -    {}^{j} \varepsilon _{n} -\mu   {\kern 1pt} \cot   (  {}^{j} \varepsilon _{n} )\right)    \left({\kern 1pt} \cot  (  {}^{j} \varepsilon _{n}   )    +    \cot^{3} (  {}^{j} \varepsilon _{n}   ){\kern 1pt} \right)}{2    \left(\mu -1+\mu \cot^{2} (  {}^{j} \varepsilon _{n} )\right)^{2} }   \right],
\end{multline}
upon substituting the function $f$ from Eq. \eqref{GrindEQ__23_} into Eq. \eqref{GrindEQ__26_}.

 In principle, using the general Householder method, formulae of order higher than 3 can be produced. The root approximation strategy proposed herein is based on successively applying Eq. \eqref{GrindEQ__21_} and Eq. \eqref{GrindEQ__25_} or Eq. \eqref{GrindEQ__27_}. These two variants lead to a second and third order method respectively. The exact steps, involved in the general approximation algorithm, are summarised in Algorithm \ref{algorithm2}.

\begin{center}
	\begin{algorithm}
		\caption{Root Approximation Strategy}
		\label{algorithm2}
		\begin{algorithmic}
			\State{Given ${}^{0} \kappa _{n} $, for $j=0,  1,  2,  ...$ }
			\State
			\State{\quad(1) Calculate ${}^{j+1} \varepsilon _{  n}       =      {\rm Arctan}\left(\mu     /{}^{j} \kappa _{n} \right)$}
			\State
			\State{\quad(2) Calculate ${}^{j  +  2} \varepsilon _{  n} $ by applying Eq. \eqref{GrindEQ__24_} or \eqref{GrindEQ__26_}, replacing $j$ by $j+1$}
			\State
			\State{\quad(3) Set $^{j  +  2} \kappa _{  n}       =      n  \pi     -    {}^{j  +  2} \varepsilon _{  n} $}
			\State{\quad(4) Replace $j+2$ by $j+1$ (in the left-hand side only) to derive the combined, two-step,\\ \quad\quad\quad iteration formula} 
			
		\end{algorithmic}
	\end{algorithm}
\end{center}

%
Applying now the above described procedure to the iteration formulae \eqref{GrindEQ__25_} and \eqref{GrindEQ__27_}, we obtain the following two iteration schemes:
\begin{itemize}
	\item \textbf{2${}^{nd}$  order method }
	\begin{equation} \label{GrindEQ__28_}
	^{j  +  1} \kappa _{  n}         =        n    \pi       +      \frac{\mu     \left(  n    \pi       -      {}^{j} \kappa _{  n}   \right)}{\mu ^{  2}       +      {}^{j} \kappa _{  n}^{  2}       -      \mu }       -      \frac{\mu ^{  2}       +      {}^{j} \kappa _{  n}^{  2} }{\mu ^{  2}       +      {}^{j} \kappa _{  n}^{  2}       -      \mu }     {\rm Arctan}    \left(\frac{\mu }{{}^{j} \kappa _{  n} } \right).
	\end{equation}
	\item \textbf{3${}^{rd}$ order method }
	\begin{equation} \label{GrindEQ__29_}
	\begin{array}{l} {^{j+1} \kappa _{  n}       =      n  \pi       +      \frac{\mu \left(  n  \pi     -    {}^{j} \kappa _{  n} \right)}{\mu ^{  2} +{}^{j} \kappa _{  n}^{  2} -\mu }       -      \frac{\mu ^{  2}     +    {}^{j} \kappa _{  n}^{  2} }{\mu ^{  2} +{}^{j} \kappa _{  n}^{  2} -\mu }     {\rm Arctan}\left(\frac{\mu }{{}^{j} \kappa _{  n} } \right)      } \\ {                                                                                                -      \frac{{}^{j} \kappa _{  n}   \mu \left(  \mu ^{  2}     +    {}^{j} \kappa _{  n}^{  2}   \right)}{\left(\mu ^{  2} +{}^{j} \kappa _{  n}^{  2} -\mu \right)^{3} }     \left[  n  \pi     -    {}^{j} \kappa _{  n}     -    {\rm Arctan}\left(\frac{\mu }{{}^{j} \kappa _{  n} } \right)\right]^{  2} .} \end{array}
	\end{equation}
\end{itemize}
The following remarks are in order at this point:
\begin{remark}
	Iteration formulae \eqref{GrindEQ__25_}, \eqref{GrindEQ__27_} and \eqref{GrindEQ__28_}, \eqref{GrindEQ__29_} are \textit{relatively simple}. Each one of them involves only a few algebraic operations and the calculation of one transcendental function.
	\end{remark}
\begin{remark}
	Setting ${}^{0} \kappa _{n} =n\pi $ in Eqs. \eqref{GrindEQ__28_}, \eqref{GrindEQ__29_}, explicit forms of reasonable accuracy are produced, which can in turn be used for the initiation of successive approximations through \eqref{GrindEQ__28_} and \eqref{GrindEQ__29_}. This direction is exploited further in the next section.
\end{remark}
\begin{remark}
	The above-described, two-step (compound) procedure, Eqs. \eqref{GrindEQ__28_} and \eqref{GrindEQ__29_}, has \textit{improved convergence characteristics} in comparison with the single-step iterations \eqref{GrindEQ__25_} and \eqref{GrindEQ__27_}. More precisely, as shown in the next proposition, although approximations \eqref{GrindEQ__28_} and \eqref{GrindEQ__29_} are of second and third order, respectively, the same as the parent single-step iterations, their error constants are significantly smaller.
\end{remark}
\begin{proposition}
	Assume that no error is introduced in all function calculations. Since, by their construction, the iteration formulae \eqref{GrindEQ__25_} and \eqref{GrindEQ__27_} are of second and third order, respectively, there exist positive constants $r,  R$, depending only on $n,  \mu $, such that (for appropriate initial guess) the iteration methods \eqref{GrindEQ__25_} and \eqref{GrindEQ__27_} converge and satisfy the error estimates
	\begin{equation} \label{GrindEQ__30_}
	\left|{}^{j+1} \varepsilon _{n} -    \varepsilon _{n} \right|      \le       r_{n} (\mu )    \left|{}^{j} \varepsilon _{n} -    \varepsilon _{n} \right|^{  2}     +      O\left(  \left|{}^{j} \varepsilon _{n} -    \varepsilon _{n} \right|  ^{3} {\kern 1pt} \right),
	\end{equation}
	\begin{equation} \label{GrindEQ__31_}
	\left|{}^{j+1} \varepsilon _{n} -    \varepsilon _{n} \right|      \le       R_{n} (\mu )    \left|{}^{j} \varepsilon _{n} -    \varepsilon _{n} \right|^{  3}     +      O\left(  \left|{}^{j} \varepsilon _{n} -    \varepsilon _{n} \right|^{  4} {\kern 1pt} \right),
	\end{equation}
	respectively. Then, the compound iterations, defined by Eqs. \eqref{GrindEQ__28_} and \eqref{GrindEQ__29_}, are also of second and third order, respectively, but now satisfy the error estimates
	\begin{equation} \label{GrindEQ__32_}
	\left|{}^{j+1} \kappa _{n} -    \kappa _{n} \right|      \le       \delta _{n}^{2} (\mu )    r_{n} (\mu )    \left|{}^{j} \kappa _{n} -    \kappa _{n} \right|^{  2}     +      O\left(  \left|{}^{j} \kappa _{n} -    \kappa _{n} \right|^{  3} {\kern 1pt} \right),
	\end{equation}
	\begin{equation} \label{GrindEQ__33_}
	\left|{}^{j+1} \kappa _{n} -    \kappa _{n} \right|      \le       \delta _{n}^{3} (\mu )    R_{n} (\mu )    \left|{}^{j} \kappa _{n} -    \kappa _{n} \right|^{  3}     +      O\left(  \left|{}^{j} \kappa _{n} -    \kappa _{n} \right|^{  4} {\kern 1pt} \right),
	\end{equation}
	respectively, where $\delta _{n} (\mu )      =      4\mu/(2n-1)^{2} \pi ^{2} +4\mu ^{2} <0.32$, for all $\mu $ and $n$. That is, the positive constants controlling the convergence rate are now considerably smaller, since they are multiplied by the small factors $\delta _{n}^{2} (\mu )$ and $\delta _{n}^{3} (\mu )$, respectively.
	\end{proposition}
The proof of the above proposition is given in Appendix \ref{sec:App}. The factor $\delta _{n}^{} (\mu )$ takes its maximum value at $\mu     =    (  n    -    1/2)  {\kern 1pt} \pi $, and satisfies the inequality
\[\delta _{n} (\mu )      =      \frac{4\mu }{(2n-1)^{2} \pi ^{2} +4\mu ^{2} }       \le       \frac{1}{(  2  n    -    1  )  {\kern 1pt} \pi }       \le       \frac{1}{\pi }       <      0.32. \]
Clearly, it decreases\textbf{ }with increasing $n$, and its value tend to zero (for all $n    {\kern 1pt} \in     {\kern 1pt} {\rm {\rm N}}$) in both limiting cases $\mu \to 0$ and $\mu \to \infty $. A plot of the function $\delta _{n} (\mu )$, for $n=1,  2,  3,  4$, is shown in Fig. 3.

 \section{Initiation, performance, and closed-form approximations}
\label{sec:5}

 Improved initiation formulae for the iterations \eqref{GrindEQ__28_} and \eqref{GrindEQ__29_} can be derived by means of Eq. \eqref{GrindEQ__21_} and the Eqs. \eqref{GrindEQ__28_} and \eqref{GrindEQ__29_} themselves. Substituting ${}^{0} \kappa _{n} =n\pi $ in Eqs. \eqref{GrindEQ__21_} and \eqref{GrindEQ__28_}, and denoting the obtained we obtain ${}^{1} \kappa _{n} $ by $A_{  n} (\mu )$ and $B_{  n} (\mu )$, respectively, we obtain the formulae

\begin{figure*}
	\begin{center}
		\includegraphics*[width=1.0\textwidth]{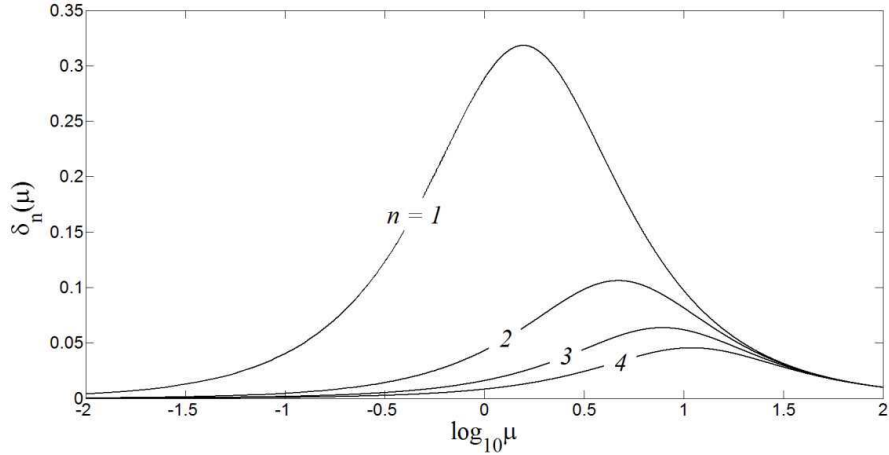}
		\caption{Plot of the factor $\delta _{n}   (  \mu   )$, as a function of parameter $\mu $, for different values of $n\in \mathbb{N}$}
		\label{fig:3}       
	\end{center}
\end{figure*}

\begin{align} 
A_{  n} (\mu )     & =      n  \pi     -    {\rm Arctan}\left(\frac{\mu }{n  \pi } \right), \label{GrindEQ__34_}\\
B_{  n} (\mu )      &=      n  \pi     -    \frac{\mu ^{2} +    n^{2} {\kern 1pt} \pi ^{2} }{\mu ^{2} +    n^{2} {\kern 1pt} \pi ^{2} -    \mu }     {\rm Arctan}\left(\frac{\mu }{n  \pi } \right). \label{GrindEQ__35_}
\end{align}
Eqs. \eqref{GrindEQ__34_} and \eqref{GrindEQ__35_} can be considered either as improved initiations of the iterations \eqref{GrindEQ__28_} and \eqref{GrindEQ__29_}, or as simple, closed-form approximations of the sought-for roots $\kappa _{  n} $. The closed-form approximation
\begin{equation} \label{GrindEQ__36_}
C_{  n} (\mu )      =      n  \pi     -    \frac{\pi }{2}     \tanh \left(\frac{2  \mu }{n  \pi ^{2} } \right),
\end{equation}
proposed by Chamberlain \cite{Ref29}, is also considered for comparison.

\begin{figure*}
	\begin{center}
	\includegraphics*[width=1\textwidth]{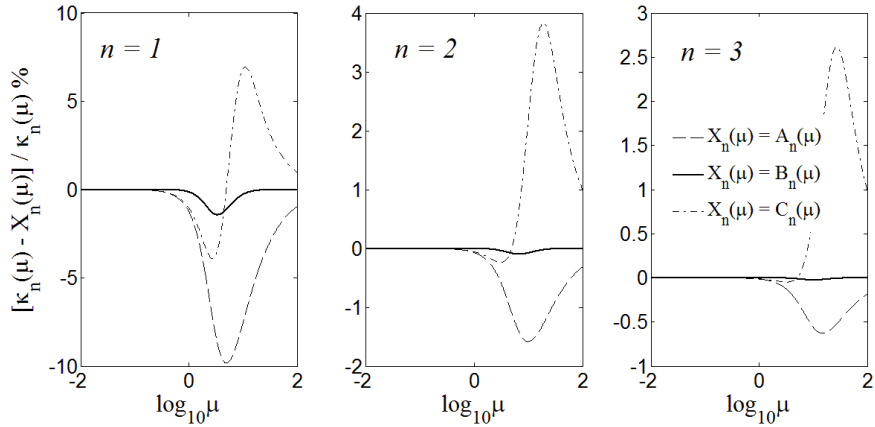}
		\caption{Relative \% error of the computation of the first three roots of Eq. \eqref{eq:knn} using $A_{\, n} (\mu )$, Eq. \eqref{GrindEQ__34_}, $B_{\, n} (\mu )$, Eq. \eqref{GrindEQ__35_}, and $C_{\, n} (\mu )$, Eq. \eqref{GrindEQ__36_}, plotted as a function of the parameter $\mu $}
		\label{fig:4}       
	\end{center}
\end{figure*}

 Numerical experiments have been performed in order to test the accuracy of the approximations \eqref{GrindEQ__34_} -- \eqref{GrindEQ__36_}. ``Exact'' solutions are obtained using the \texttt{fzero} function of \textsc{Matlab} with initial values given by Eq. \eqref{GrindEQ__35_}. The percentage of relative error of $A_{  n} (\mu )  ,  {\kern 1pt} B_{  n} (\mu )  ,  {\kern 1pt} C_{  n} (\mu )$, as function of $\mu $, is plotted in Figure \ref{fig:4}, for the first three values of $n$. For $A_{  n} (\mu )$ and $C_{  n} (\mu )$ the maximum relative error is less than 10\%, while for $B_{n} (\mu )$ the maximum relative error is less than 1.5\%. The error drops significantly as $n$ increases. In all cases, the expression $B_{n} (\mu )$ performs better than $A_{  n} (\mu )$ and $C_{  n} (\mu )$.
 
\begin{figure*}
	\begin{center}
			\includegraphics*[width=1.0\textwidth]{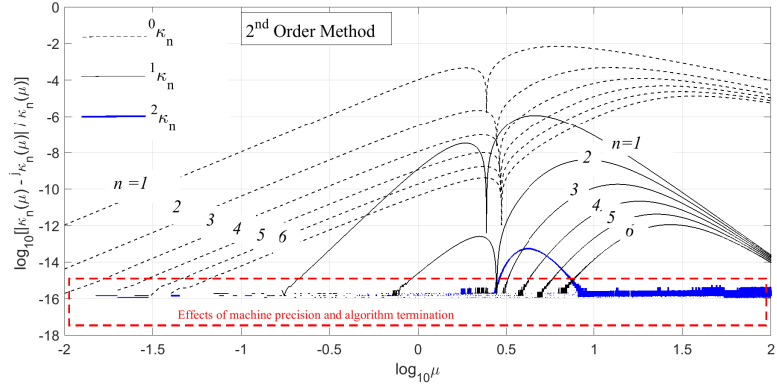}
			\caption{Logarithmic plot of the relative error of the approximation $D_{  n}   (  \mu   )$ and the two first iterations of Eq. (28) initiated by $D_{  n}   (  \mu   )$ for the first 6 roots of Eq. (16)}
			\label{fig:5}       
	\end{center}
\end{figure*}

 In accordance to the procedure described in Algorithm \ref{algorithm2}, the quantities $n  \pi $, $A_{  n}   (  \mu   )$ and $B_{  n}   (  \mu   )$ are three consecutive terms of a sequence converging to $\kappa _{  n} $. Motivated by this fact, we apply a Shanks' transformation \cite[p. 369]{Ref64}, also known as Aitken $\delta ^{  2} -$method or Steffensen's procedure \cite[p. 103]{Ref65}, to produce a possibly better approximation:
\begin{equation} \label{GrindEQ__37_}
D_{n} (\mu )      =      n  \pi       -      \frac{\left[A_{\, n} (\mu )\, \, -\, \, n\, \pi \right]^{\, 2} }{\left[n\, \pi \, \, +\, \, B_{\, n} (\mu )\, \, -\, \, 2A_{\, n} (\mu )\right]} .
\end{equation}
Substituting $A_{\, n} (\mu )$ and $B_{\, n} (\mu )$ from Eqs. \eqref{GrindEQ__34_}, \eqref{GrindEQ__35_} into Eq. \eqref{GrindEQ__37_}, we obtain
\begin{equation} \label{GrindEQ__38_}
D_{\, n} (\mu )\, \, \, =\, \, \, n\, \pi \, \, -\, \, \left[\frac{\mu ^{2} +\, \, n^{2} {\kern 1pt} \pi ^{2} -\, \, \mu }{\mu ^{2} +\, \, n^{2} {\kern 1pt} \pi ^{2} -\, \, 2\, \mu } \right]\, \, {\rm Arctan}\left(\frac{\mu }{n\, \pi } \right).
\end{equation}
Formula \eqref{GrindEQ__38_} is ``almost the same'' as formula \eqref{GrindEQ__35_}, and it is indeed more accurate than the latter. The maximum relative error in the computation of the first root is about 0.7\%; see Figure \ref{fig:5}. Again, the maximum relative error reduces significantly for increasing $n$.

\begin{figure*}
	\begin{center}
		\includegraphics*[width=1.0\textwidth]{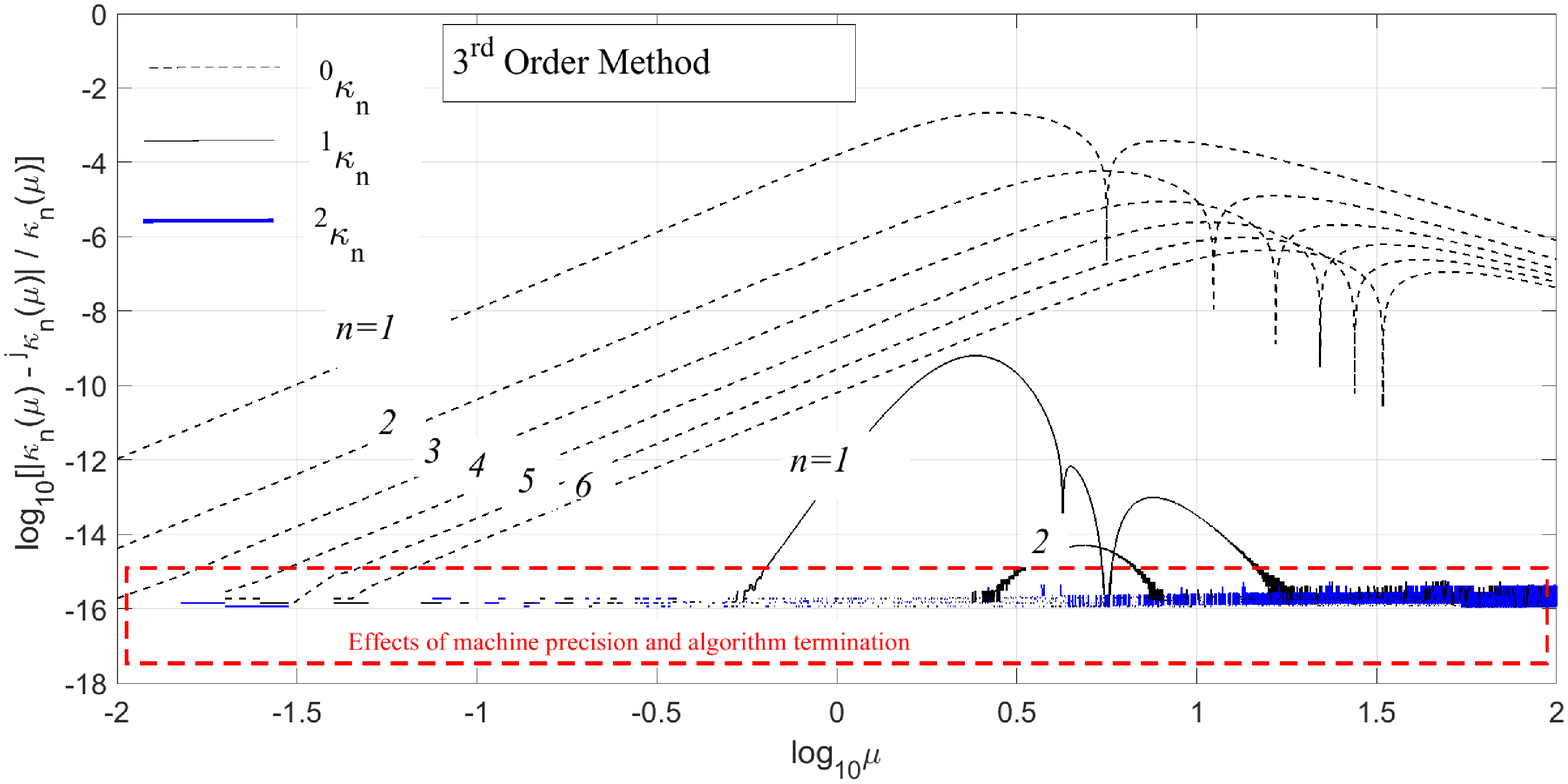}
		\caption{Logarithmic plot of the relative error of the approximation $E_{ n}  ( \mu  )$ and the two first iterations of Eq. (29) initiated by $E_{ n}  ( \mu  )$ for the first 6 roots of Eq. (16)}
		\label{fig:6}       
	\end{center}
\end{figure*}

 Finally, a different approximation is produced by setting ${}^{0} \kappa _{n} =n\pi $ in the third-order formula \eqref{GrindEQ__29_}, obtaining:
\begin{multline} \label{GrindEQ__39_}
E_{n} (\mu )=n\pi -\frac{\mu ^{2} +n^{2} \pi ^{2} }{\mu ^{2} +n^{2} \pi ^{2} -\mu } {\rm Arctan}\left(\frac{\mu }{n\pi } \right)\\
-\frac{n\pi \mu \left(\mu ^{2} +n^{2} \pi ^{2} \right)}{\left(\mu ^{2} +n^{2} \pi ^{2} -\mu \right)^{3} } \left[{\rm Arctan}\left(\frac{\mu }{n\pi } \right)\right]^{2} .
\end{multline}
The error of $E_{n} (\mu )$ is plotted in Figure  \ref{fig:6}. Comparing Figures \ref{fig:5} and \ref{fig:6}, it is readily seen that $E_{n} (\mu )$ behaves better that $D_{n} (\mu )$ for large values of $\mu $.

 Approximations $D_{n} (\mu )$ and $E_{n} (\mu )$ can be used for the initiation of the iteration formulae \eqref{GrindEQ__28_} and \eqref{GrindEQ__29_}, respectively. As shown in Figures \ref{fig:5} and \ref{fig:6}, they produce highly accurate results after the first iteration and, in general, results to machine precision after the second iteration. Especially, Eq. \eqref{GrindEQ__29_} initiated by $E_{n} (\mu )$, leads to the determination of $\kappa _{ n}  ( \mu  )$ with an error less than $10^{ -  15} $ for the whole range of $\mu $. Both methods appear to yield exact results for a particular value of $\mu $, that depends on the number of the mode under consideration. This value of $\mu $ tends to be higher, as the number of mode increases.

\begin{figure*}
	\begin{center}
		\includegraphics*[width=1.0\textwidth]{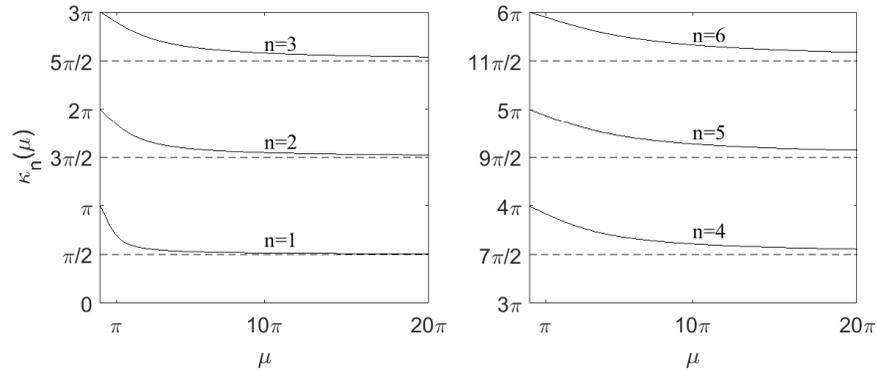}
		\caption{Plot of the first six roots of Eq. (16), as approximated by Eq. (29), initiated by $E_{ n}  ( \mu  )$, with two iterations}
		\label{fig:7}       
	\end{center}
\end{figure*}

 The main advantage of the present schemes in comparison with those proposed by Newman \cite{Ref27}, is the presence of powers of the small factor $\delta _{n} (\mu )$ in the error constant (see Proposition 1 and Remark 3 above). It is remarkable that the regions of increased errors in Fig. 5 and 6 coincide, more or less, with the region of high values $\delta _{n} (\mu )$, as shown in Figure \ref{fig:3}. In any case, these errors are already small and disappear at the second iteration. The first six roots $\kappa _{\, n} $, as obtained by two iterations of Eq. \eqref{GrindEQ__29_}, initiated by $E_{\, n} \, (\mu )$, are plotted in Figure \ref{fig:7}, as functions of $\mu $.

\begin{figure*}
	\begin{center}
		\includegraphics*[width=1.0\textwidth]{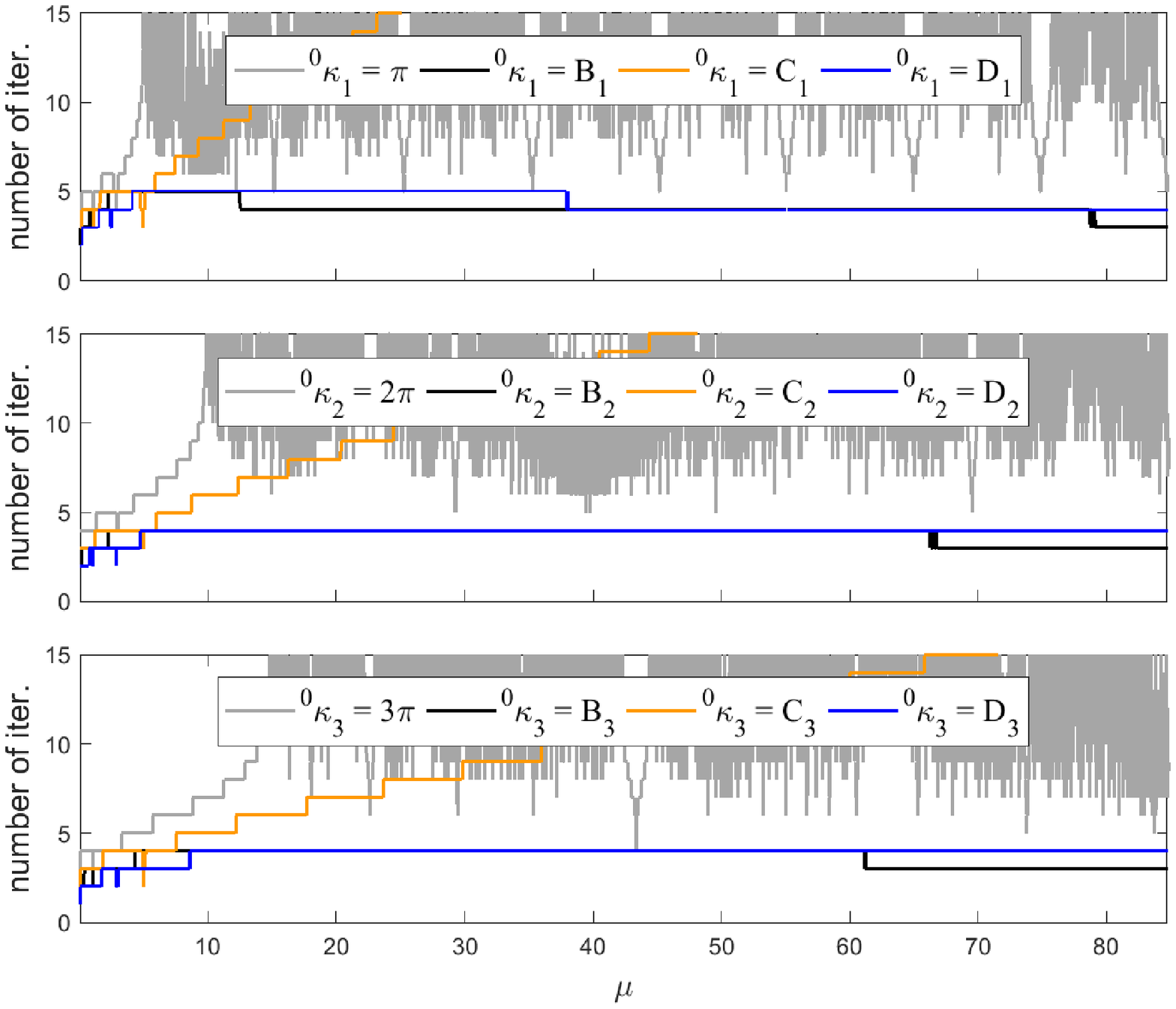}
		\caption{Number of iterations of the Newton-Raphson method for the first three roots of Eq. \eqref{GrindEQ__16_}, for initial guesses $^{0} \kappa _{n}     =    n\pi   ,  {\kern 1pt}   B_{  n}   (\mu )  ,  {\kern 1pt}   C_{  n}   (\mu )  ,  {\kern 1pt}   D_{  n}   (\mu )$}
		\label{fig:8}       
	\end{center}
\end{figure*}

 Before proceeding with the application of the above results to the simulation of demanding, nonlinear, water-wave problems, we shall discuss the enhancement of the Newton-Raphson (NR) method by the exploitation of Eq. \eqref{GrindEQ__35_} - \eqref{GrindEQ__37_} for its initialization. For $\mu     {\kern 1pt} \in     {\kern 1pt} (  0  ,  {\kern 1pt} 85  )$, we compare the number of iterations required by the NR method in order to reach machine precision tolerance by starting from the following four different initial guesses: $^{0} \kappa _{n}     =    n\pi   ,  {\kern 1pt}   B_{  n}   (\mu )  ,  {\kern 1pt}   C_{  n}   (\mu )  ,  {\kern 1pt}   D_{  n}   (\mu )$. Results for the first three roots, $\kappa _{  1}   ,  {\kern 1pt} \kappa _{  2}   ,  {\kern 1pt} \kappa _{  3} $, are shown in Figure \ref{fig:8}. As one might expect, the choice $^{0} \kappa _{n}     =    n\pi $ is sufficient only when $\mu $ is relatively small. The number of iterations increases rapidly with increasing $\mu $, leading eventually to overshooting or divergence of the NR method after some value $\mu _{  *} $, which increases with the root index $n$. The initial guess $^{0} \kappa _{n}     =    C_{n} $ leads to convergent NR iterations, but the number of iterations increases significantly for large$\mu $. On the other hand, the choices $^{0} \kappa _{n}     =    B_{n}   ,  {\kern 1pt} D_{n} $ are the more robust, leading to convergence for the whole range of $\mu $ after $4-5$ NR iterations.

\section{Applications}
 \label{sec:6}
In this section, we present a number of applications of the HCMT to specific, nonlinear, water-wave problems, and discuss in detail the role of the new approximation formulae for the roots $\kappa _{  n} $ to the implementation of the method. Before proceeding to the specific applications, it is expedient to recall that the HCMS Eqs. \eqref{GrindEQ__7_}, closed by the substrate kinematical problem \eqref{GrindEQ__8_}, is just an exact reformulation of the fully nonlinear water-wave problem, Eqs. \eqref{GrindEQ__1_}, \eqref{GrindEQ__2_}, provided that the coefficients $A_{  m  n} $, $B_{  m  n} $ and $C_{  m  n} $ are exactly calculated. Given that $A_{  m  n} $, $B_{  m  n} $ and $C_{  m  n} $ have been analytically calculated in terms of $\kappa _{  n} $, the issue of the accuracy of the new Hamiltonian formulation is fully controlled by the accuracy of the determination of $\kappa _{  n} $. Having ensured the latter, by using the results of Sections \ref{sec:4} and \ref{sec:5}, the new HCMS becomes very efficient because of the \textit{striking superconvergence} of $\varphi _{  -    2} $ with respect to the number of modes, $N_{  {\rm tot}} $, retained in the truncated version of system \eqref{GrindEQ__8_}, as already mentioned. Thus, only a few modes (up to five or six) are enough for accurate hydrodynamic calculations.

 In problems studied in this section, the initialization of the evolutionary Hamiltonian equations \eqref{GrindEQ__7_} is performed by means of an accurately pre-calculated steady travelling wave (in constant depth), of specific wavelength and height. Such a wave provides the initial conditions in the case studied in Subsection \ref{subsec:61}, while it plays the role of the excitation (incoming wave) in the cases studied in Sections \ref{subsec:62} and \ref{subsec:63}. High-accuracy numerical simulations of steady travelling waves can be obtained by various methods/codes provided, for example, by \cite{Ref65}, \cite{Ref66}, \cite{Ref67} or an appropriate version of the present HCMT, given in \cite[Chapter 6]{Ref59}. The latter method was used in all cases presented below. Note that all methods give practically identical results.
 
 \subsection{Evolution of steady travelling waves over flat bottom}
 \label{subsec:61}
  The problems considered in this subsection are simple and permit an easy monitoring of the accuracy of the numerical solution. HCMS, Eqs. \eqref{GrindEQ__7_}, is solved under periodic lateral conditions, initialized by the fields $(  \eta   (  x  ,  t_{  0}   )  ,   \psi   (  x  ,   t_{ 0}  ) ) = (\eta _{ 0}  ( x ) ,  \psi_{0}  ( x ) )$ corresponding to a travelling nonlinear waver with wave length equal to the horizontal extent of the domain. Thus, the evolution equations have to reproduce the initial state after one period, two periods etc. permitting a trivial accuracy check. Three different waves are considered, corresponding to deep, intermediate and shallow water conditions. Normalizing the depth to $h_{  0}     =    1$, the wave lengths of the three cases are chosen to be $\lambda     =    1$ m (deep water), $\lambda     =    5$ m (intermediate-depth water), and $\lambda     =    18$ m (shallow water). In all cases, the wave height has been selected to be the 80\% of the maximum value for the specific choice of $\lambda /h$, as predicted by Williams \cite{Ref68}. The parameter $\mu _{{\kern 1pt} 0} $ is chosen to be $\mu _{{\kern 1pt} 0}     =    (2\pi /\lambda )  {\kern 1pt} \tanh (  2\pi   h/\lambda   )$, that is, the value of $\mu _{  0} $ corresponding to the linear waves with the same wavelength. This is the most appropriate choice in this case, since it ensures that the representation of the flow field by the series \eqref{GrindEQ__3_} already encapsulates the physics of the corresponding linear wave, even before starting the procedure of the numerical solution. Of course, to obtain an accurate approximation, Eq. \eqref{GrindEQ__16_} is solved for the varying $\mu   (  x  )    =$ $\mu _{{\kern 1pt} 0}   (  \eta   (  x  )  +  {\kern 1pt} h  )$, so that the local vertical basis, at each $(  x  ,  {\kern 1pt} t  )$, to accurately represent the local wave potential field.

\begin{figure*}
 	\begin{center}
 		\includegraphics*[width=1.0\textwidth]{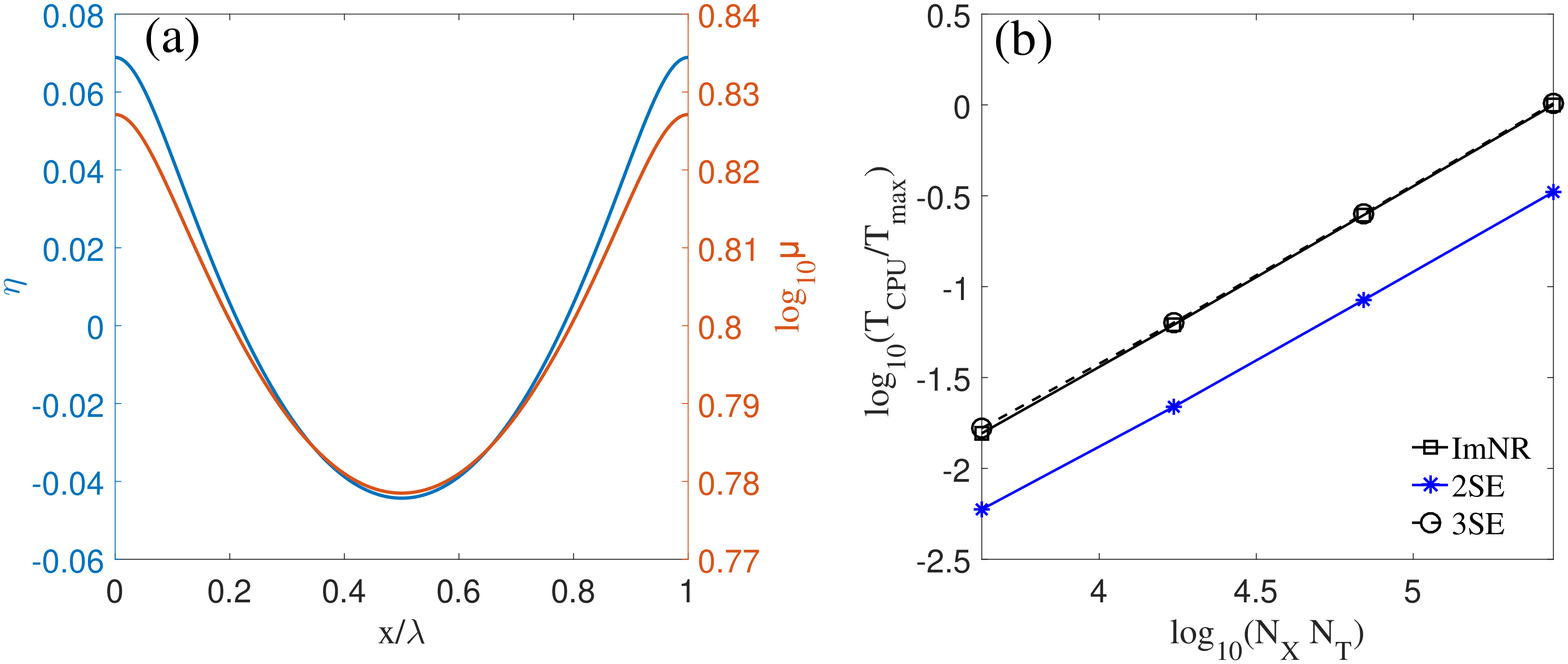}
 		\caption{ (a) Free surface elevation $\eta  ( x )$ and $\log _{10} \mu ( x )$ for the case $\lambda /h = 1$. (b) Relative computational time for the evaluation of $\kappa _{n} $ versus $N_{X} N_{T} $ for the improved NR and the semi-explicit methods}
 		\label{fig:9}       
 	\end{center}
 \end{figure*}

\begin{figure*}
 	\begin{center}
 		\includegraphics*[width=1.0\textwidth]{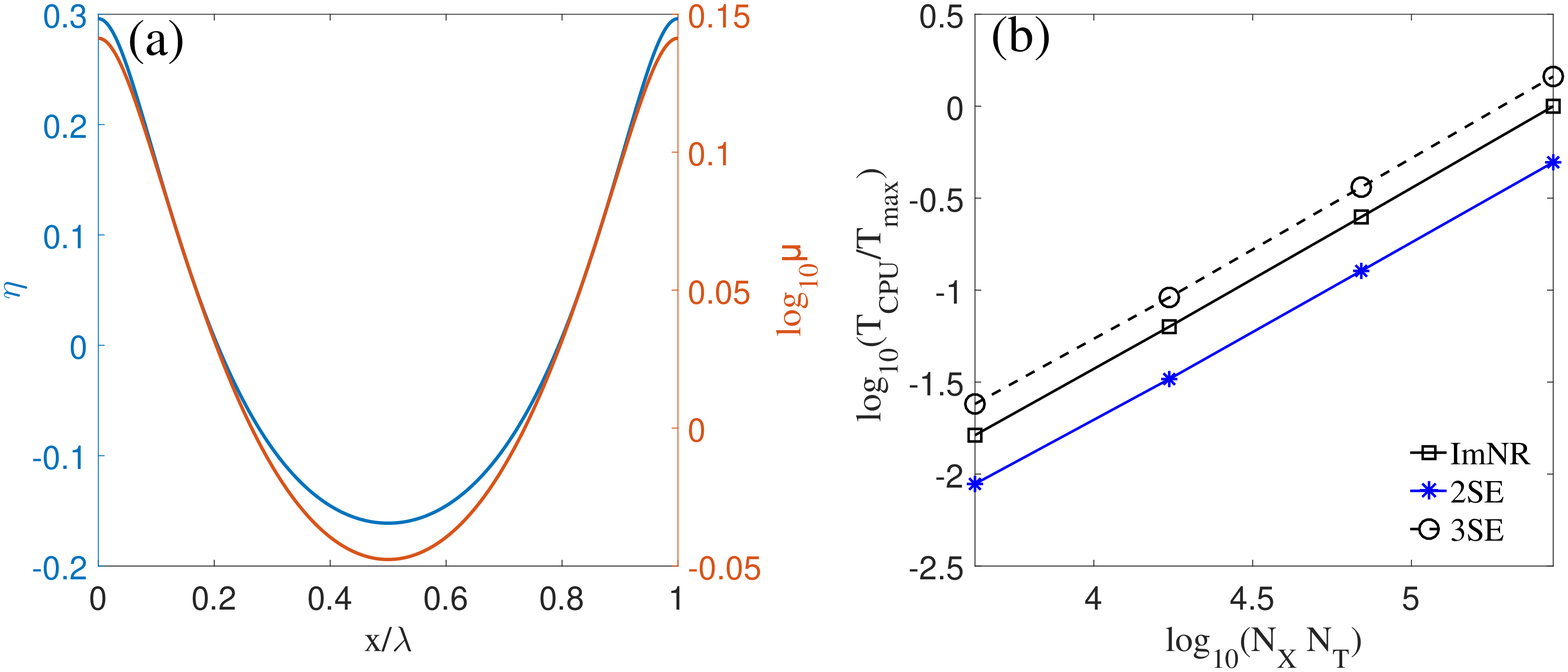}
 		\caption{Same as Figure 9 for the case $\lambda /h = 5$}
 		\label{fig:10}       
 	\end{center}
 \end{figure*}

\begin{figure*}
	\begin{center}
		\includegraphics*[width=1.0\textwidth]{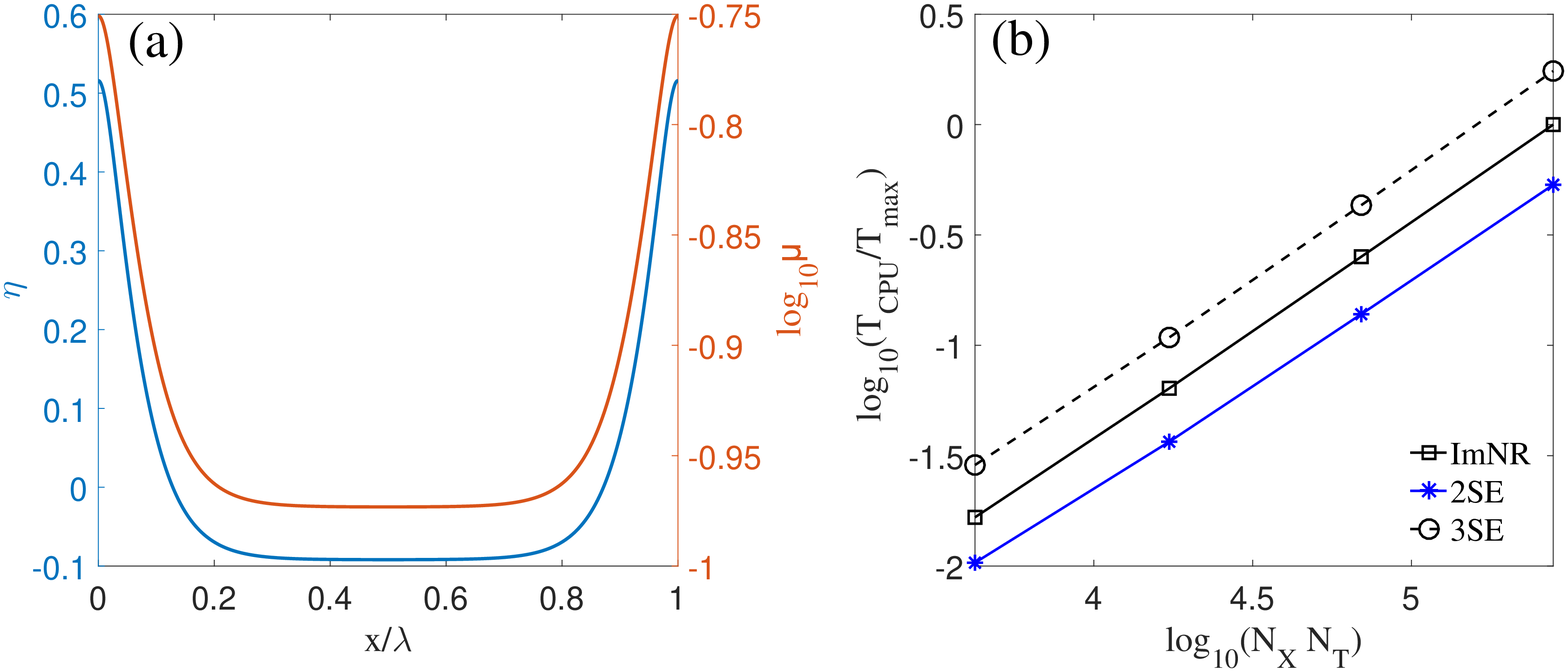}
		\caption{ Same as Figure \ref{fig:9} for the case $\lambda /h = 18$}
		\label{fig:11}       
	\end{center}
\end{figure*}

 The free-surface wave profiles for the three cases studied are shown in the left panels of Figures \ref{fig:9}-\ref{fig:11}, along with the corresponding local values of $\mu     =    \mu (  x  )    =    \mu _{{\kern 1pt} 0}   (  \eta (  x  )  +  {\kern 1pt} h  )$. The evolution problems are solved for a time span of three periods, by using $N_{{\rm tot}}     =    3  {\kern 1pt} ,  {\kern 1pt} 4  {\kern 1pt} ,  {\kern 1pt} 5  ,  {\kern 1pt} 6$ modes, that is, the modes $-2  ,  {\kern 1pt} -1$, supplemented by 1, 2, 3, 4 evanescent modes. The coefficients $A_{  m  n} $, $B_{  m  n} $, $C_{  m  n} $ are calculated by means of closed-form equations, like Eqs. \eqref{eq:ABCsAn}, in terms of the roots $\kappa _{  n} $ of Eq. \eqref{GrindEQ__16_}. The whole numerical scheme is implemented as explained in Subection. 3.1.

 For comparison purposes, Eq. \eqref{GrindEQ__16_} is solved by four different methods:

\begin{enumerate}
\item  The standard Newton -- Raphson (\textbf{StNR}) method applied to Eq. \eqref{GrindEQ__16_}, initialized by ${}^{0} \kappa _{  n}     =    n  \pi $, with tolerance $10^{-15} $ (with as many iterations $J_{  {\rm iter}}   (  n  )$ as needed),

\item  The NR method applied to Eq. \eqref{GrindEQ__16_}, now initialized by $B_{  n}   (  \mu   )$, Eq. \eqref{GrindEQ__35_}, with the same tolerance $10^{-15} $, called subsequently the improved NR (\textbf{ImNR}) method,

\item  The 2${}^{nd}$ order semi-explicit (\textbf{2SE}) method, Eq. \eqref{GrindEQ__28_}, initiated by $D_{  n}   (  \mu   )$, Eq. \eqref{GrindEQ__38_}, with $J_{  {\rm iter}}   (  1  )    =    3$ and $J_{  {\rm iter}}   (  n  )    =    2$, for $n    \ge     2$,

\item  The 3${}^{r}$${}^{d}$ order semi-explicit (\textbf{3SE}) method, Eq. \eqref{GrindEQ__29_}, initiated by $E_{  n}   (  \mu   )$, Eq. \eqref{GrindEQ__39_}, with $J_{  {\rm iter}}   (  1  )    =    J_{  {\rm iter}}   (  2  )    =    2$, and $J_{  {\rm iter}}   (  n  )    =    1$ for $n    \ge     3$.
\end{enumerate}

 A first important numerical finding from the applications presented below, is that the StNR method does not converge for $\kappa _{  1} $ in the deep water case ($\lambda /h    =    1$), since the values of $\mu     =    \mu   (  x  )$ become large (of the order of 10). The ImNR method, however, does converge in all studied cases. The two semi-explicit methods work with pre-specified number of iterations, being thus free of the convergence question.
 
 In order to illustrate the nonlinear accuracy and efficiency of the present method we investigate the relative $L^{2} -{\rm error}$ of the free surface elevation, computed by (cf. \cite{Ref10})
\begin{equation*}
\text{Error} = \frac{||\eta _{  3}     -    \eta _{  0} ||_{  2} }{3    ||\eta _{  0} ||_{  2} }
\end{equation*}
where $\eta _{  {\rm 0}} $ is the given initial wave profile, and $\eta _{  3} $ is the computed wave profile after three periods. The horizontal grid consists of $N_{X}     =    12  8$ points in the cases $\lambda /h    =    1  {\kern 1pt} ,  {\kern 1pt} 5$, and $N_{X}     =    256$ points in the case $\lambda /h    =    18$. The Courant--Friedrichs--Lewy (CFL) number, based on the group velocity of the nonlinear wave, is 0.7 for all cases. The roots $\kappa _{  n} $ of Eq. \eqref{GrindEQ__16_} are computed by using the 2SE method. Results concerning the error are shown in Table \ref{tab:1}. The convergence of the numerical scheme with respect to the number of modes $N_{{\rm tot}} $ is clearly demonstrated in the same table. It is clearly seen that a total number of 5-6 modes is sufficient for an error of order $10^{-4} $-$10^{-5} $, which is comparable with the corresponding results of Bingham \& Zhang \cite[Section 5]{Ref10}, obtained by using $4^{th}$ order finite differences in the whole domain. The errors typically increase for larger simulation times and/or higher nonlinearity, which is expected in the context of the present explicit time-integration scheme. With the same spatio-temporal discretisation, a total number of modes $N_{{\rm tot}}     =    7$ was needed for a stable simulation of duration $50  T$ in the deep water case producing an error of $1.9\times 10^{-  3} $. For the same duration, the errors in the intermediate and shallow water cases (with $N_{{\rm tot}}     =    6$) are $1.0\times 10^{-  4} $ and $1.2\times 10^{-4} $. For higher nonlinearity (wave height 90\% of the maximum value) the errors after $50  T$ for the deep, intermediate and shallow case are $8.0\times 10^{-  3} $, $8.8\times 10^{-  3} $ and $3.1\times 10^{-  2} $ respectively, by using $N_{tot}     =    7$. The above computations have been performed without the use of smoothing or filtering. The complete investigation of the numerical stability and limitations of the present scheme in conjunction with steady periodic waves will be the subject of another work. In the case of solitary waves, results can be found in \cite{Ref57}.

 The relative computational time needed for the evaluation of $\kappa _{n} $, $n    =    1,  {\kern 1pt} ...  {\kern 1pt} ,  {\kern 1pt} 4$ ($N_{tot} =6$) during the simulation of three periods, is documented for increasing spatial resolution $N_{X}     =    32  {\kern 1pt} ,  {\kern 1pt} 64$, $128$, $256$. Results for the above three cases are shown in the right panels of Figures \ref{fig:9}-\ref{fig:11}. The ImNR method and the 3SE method are the most expensive in all three cases. This is due to the relatively large number of iterations as concerns the ImNR, and to the more involved mathematical expressions as regards to 3SE. Thus, 2SE method turns out to be the best choice, being always the faster one, predominantly in deep water conditions. It should be stressed once again that the standard NR method, initialized by obvious choice ${}^{0} \kappa _{  n}     =    n  \pi $, fails to converge in many cases, especially for deep water conditions. The improved NR method (that is, the NR method initiated by $B_{  n}   (  \mu   )$ or $D_{  n}   (  \mu   )$) is proven to be convergent for all values of ${\it n}$ and ${\it \mu }  {\rm (}  {\rm x}  {\rm )}  {\rm .}$ These findings underline the significance of the proposed root approximation strategy and the explicit formulae given in Section \ref{sec:5}.
\begin{table}[h]
	\caption{$L^{2} -{\rm error}$ on the free surface elevation for the simulation of strongly nonlinear travelling waves over flat bottom}
	\label{tab:1}       
	\begin{center}
	\begin{tabular}{lllll}
\hline\noalign{\smallskip}	
 & \multicolumn{4}{l}{\centering{\quad\quad\quad\quad\quad\quad${\rm E} {\rm rror}    {\rm at}    t=3T$}} \\ \hline
$\lambda /h$ & $N_{{\rm tot}}     =    3$ & $N_{{\rm tot}}     =    4$ & $N_{{\rm tot}}     =    5$ & $N_{{\rm tot}}     =    6$ \\ \hline
1 & (*) & $6.0\times 10^{-3} $ & $1.3\times 10^{-3} $ & $1.9\times 10^{-4} $ \\ \hline
5 & $4.1\times 10^{-3} $ & $3.6\times 10^{-4} $ & $4.6\times 10^{-5} $ & $9.1\times 10^{-5} $ \\ \hline
18 & $6.2\times 10^{-3} $ & $3.3\times 10^{-4} $ & $1.8\times 10^{-4} $ & $2.6\times 10^{-4} $ \\ \hline
\end{tabular}
\end{center}
 (*)  In this case, the number of modes ($N_{{\rm tot}}     =    3$) is not sufficient for the convergence of the algorithm.
\end{table}

 \subsection{Transformation of incident waves over a submerged bar }
\label{subsec:62}

 In order to further investigate the effect of semi-explicit methods derived in this paper, we consider the transformation of incident regular waves due to a submerged trapezoidal bar. This configuration has been investigated in the experiments of Beji \& Battjes \cite{Ref41} and Dingemans \cite{Ref42}, and is considered a standard benchmark test for the ability of numerical models to correctly simulate nonlinear and dispersive waves over variable bathymetry. A detailed investigation of the performance of
 
\begin{figure*}
	\begin{center}
		\includegraphics*[width=1.0\textwidth]{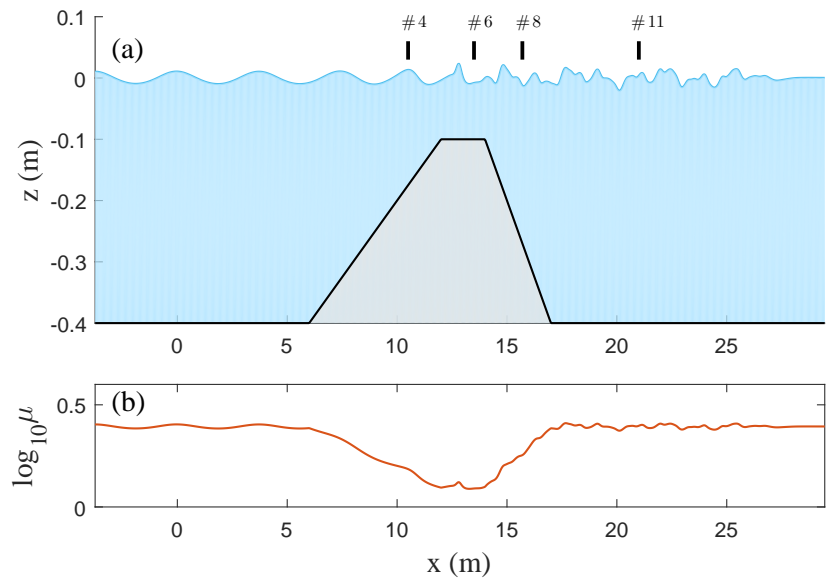}
		\caption{ (a) Instantaneous free surface, bottom surface and locations of measuring gauges in the experiment of \cite{Ref42}. (b) Values $\mu ( x , t ) = \mu _{0} ( \eta ( x ,t ) + h ( x ) )$ at the final simulation time}
		\label{fig:12}       
	\end{center}
\end{figure*}

\begin{figure*}
	\begin{center}
		\includegraphics*[width=1.0\textwidth]{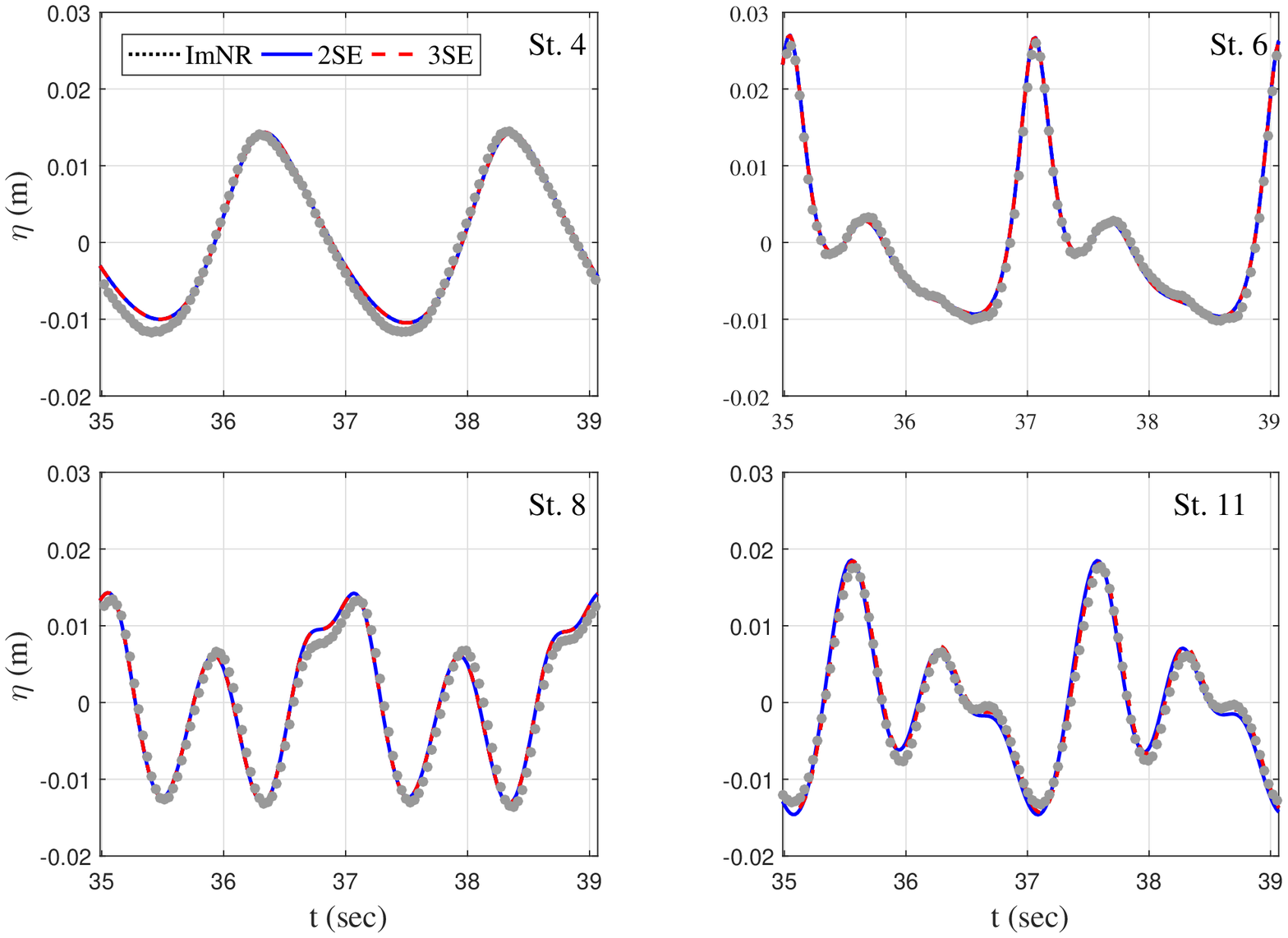}
		\caption{ Comparison of the computed free surface elevation with experimental data (grey bullets) of \cite{Ref42}}
		\label{fig:13}       
	\end{center}
\end{figure*}

HCMS in this experiment is presented in \cite{Ref16}. The results obtained here correspond to $N_{{\rm tot}}  = 7$ modes (the modes -2, -1, the propagating one, and four evanescent ones). The parameter $\mu _{ 0} $ is chosen as $\mu _{ 0}  = \omega _{ 0}^{ 2} /g$, where $\omega _{ 0}^{} $ is the circular frequency of the incident wave, $\omega _{ 0}^{}  = 2\pi /T$, with $T = 2.02$ sec. The spatio-temporal discretization is $N_{X} = 707$ and $N_{T} = 3500$. We consider the three different methods, ImNR, 2SE and 3SE, for calculating the four roots $k_{n} ,  n=1,2,3,4$, at any horizontal position $x$ and time $t$. In all three simulations, the matrix coefficients are evaluated at machine precision. The fluid domain at a specific simulation time is shown in Figure \ref{fig:12}, together with the corresponding values of $\mu  ( x )$. The validity of the simulations is illustrated in Figure \ref{fig:13} by comparison of the time series of the free surface elevation at four indicative measuring stations (St. 4,6,8,11) used in \cite{Ref42}. As expected, the computational results are identical and in very good agreement with experimental measurements. Differences are present only in the total simulation time. ImNR and 3SE methods result in practically the same computational time, while the 2SE method leads to 2\% faster simulation. The small time enhancement observed is due to the fact that the time needed for the solution of the transcendental equation \eqref{GrindEQ__16_} is small in comparison with the time needed for the numerical solution of the HCMS, Eqs. \eqref{GrindEQ__7_}, \eqref{GrindEQ__8_}. Most important, however, is the robustness of the method.

\subsection{Bragg scattering over sinusoidal bottom ripple patch}
 \label{subsec:63}
\begin{figure*}
 	\begin{center}
 		\includegraphics*[width=1.0\textwidth]{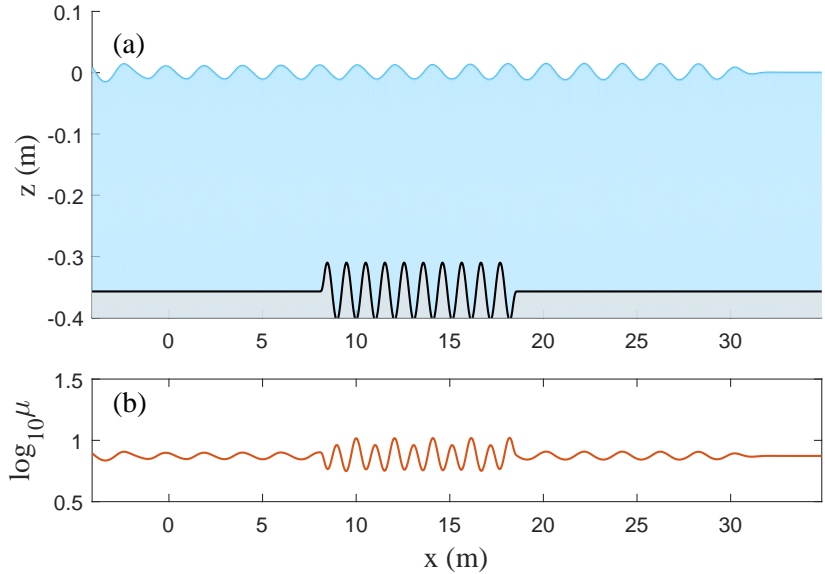}
 		\caption{(a) Instantaneous free surface and bottom surface in the experiment of [43]. (b) Values $\mu (  x  ,  t  )    =    \mu _{0} (  \eta (  x  ,t  )    +    h(  x  )  )$ at the final simulation time}
 		\label{fig:14}       
 	\end{center}
 \end{figure*}

\begin{figure*}
	\begin{center}
		\includegraphics*[scale=0.8]{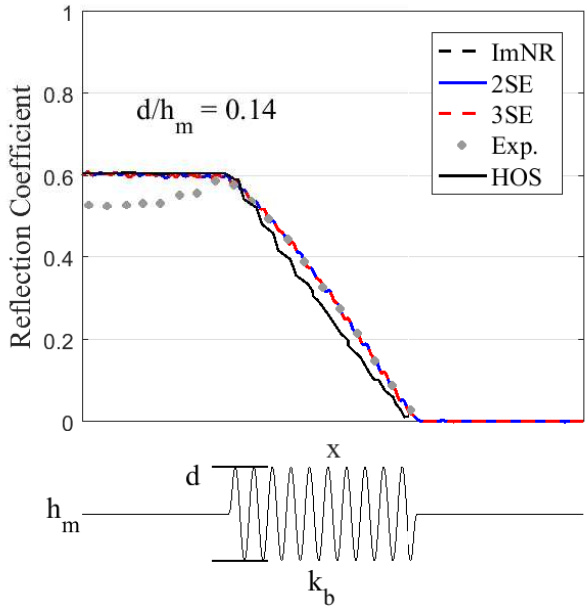}
		\caption{ Local Bragg reflection coefficient in the experiment of \cite{Ref43}}
		\label{fig:15}       
	\end{center}
\end{figure*}

 In this example, we study the transformation of an incident regular periodic wave propagating in a region having a sinusoidal patch in the seabed. We consider the configuration studied by Davies \& Heathershaw \cite{Ref43}, where this wave-bottom interaction was first investigated. The bottom sinusoidal patch, of wavelength $l_{b} $, extends from $x_{l} $ to $x_{r}     =    x_{l}     +    10  l_{b} $, and is defined by
 \begin{equation*}
 z    =    \left\{\begin{array}{c} {-  h_{m}     +    d\sin   (  k_{b}   x  )  ,\quad    x_{l}     <    x    <    x_{  r}   ,} \\ {-  h_{m} ,        {\rm elsewhere}  {\rm ,}} \end{array}\right.
 \end{equation*}
 where $k_{b}     =    2\pi /l_{b} $. The incident wave is chosen to have a wavenumber $k    =    k_{b} /2$ (Bragg resonance condition), steepness $k  H/2    =    0.05$ and period $T    =    1.28$ sec. We consider HCMS with $N_{{\rm tot}}     =    6$, $N_{X}     =    1000$, and $\mu _{  0}     =    k  \tanh   (  k  h_{  m} )$; see Figure \ref{fig:14}. Simulations correspond to a duration of $40T$, which is sufficient in order to estimate the reflection coefficient from the analysis of the local time series of the free surface elevation \cite{Ref69}. A comparison of experimental data and our computations, obtained by using the three methods for the computation of $k_{n} ,    n=1,2,3$, are shown in Figure \ref{fig:15}. Results corresponding to the Higher Order Spectral (HOS) method \cite{Ref70} have been digitized and are also shown in the same figure. Our three variants of computations are indistinguishable, as expected, and show excellent agreement with the measurements of \cite{Ref42} over the rippled patch. HOS method slightly underestimates the reflection coefficient over this region. On the upwave side, all numerical computations are in agreement, differing from the measurements of \cite{Ref42}. This is due to unwanted interference effects occurring in the experimental tank, as reported in \cite{Ref42}. As in the previous case, the total computational time is approximately the same when using ImNR and 3SE methods, being 3\% larger than the one corresponding to using 2SE method. Again, the small time enhancement is due to the fact that the time needed for the solution of Eq. \eqref{GrindEQ__16_} is small in comparison with the total time needed for the numerical solution of Eqs. \eqref{GrindEQ__7_}, \eqref{GrindEQ__8_}.

 \subsection{Disintegration of a solitary wave due to an abrupt deepening}
 \label{subsec:64}
\begin{figure*}[h]
 	\begin{center}
 		\includegraphics*[width=1.0\textwidth]{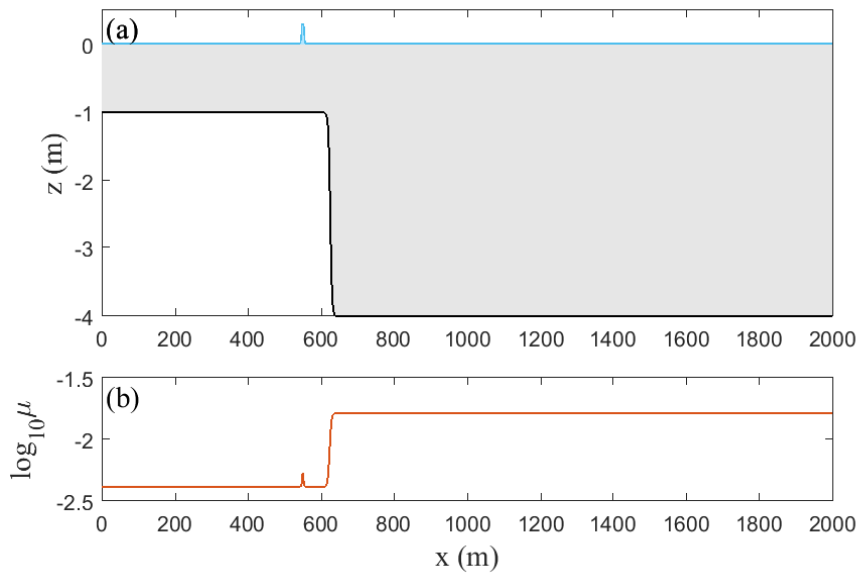}
 		\caption{(a) Initial free surface, and bottom surface for the simulation of a solitary wave propagating over an abrupt deepening, with $h_{  r}     =    4  m$. (b) Values of $\mu   (  x  ,  t  )    =$$\mu _{  0}   (  \eta   (  x  ,t  )    +    h  (  x  )  )$ at the initial time, for the same case}
 		\label{fig:16}       
 	\end{center}
 \end{figure*}

 As a final example, we consider the transformation of a solitary wave propagating over an abrupt deepening, that is, a strong depth increase. The interesting feature of this numerical experiment is the transition from shallow to not-shallow water conditions which induces multiple space and time scales, possibly out of the reach of simplified models. To the best of our knowledge, this nonlinear phenomenon has not been previously examined.
 In the numerical simulation of the above described problem, the horizontal domain extents from $x=0$ to $x=2000$m. The bottom surface is defined by the equation
 \begin{equation*}
 h  (  x  )      =      -    a    {\kern 1pt} -{\kern 1pt}     b  {\kern 1pt} \tanh   (  0.2  (  x    -  625  )  ),
 \end{equation*}
where $a    >    b    >    0$ are shape parameters defining the left and right (small and large, respectively, see Figure \ref{fig:16}) depth of the seabed:
 \begin{equation*}
h_{  l}       =      a-b  ,    h_{ r}  = a+b,
 \end{equation*}
This bathymetry represents a smooth but abrupt depth transition from $h_{  l}     =    h_{  0}     =    1    {\rm m}$ to a larger depth $h_{  r} $, which has been taken to be $h_{  r}     =    2  h_{  0}   ,    4  h_{  0}   ,    8  h_{  0} $. The initial solitary wave is centered at $x=550$ m having an amplitude $a    =    0.3$m (see Figure \ref{fig:16} (a)). Initial conditions $(  \eta   (  x  ,  {\kern 1pt} t_{  0}   )  ,  {\kern 1pt} \psi   (  x  ,  {\kern 1pt} t_{  0}   )  )    =    (  \eta _{  0}   (  x  )  ,  {\kern 1pt} \psi _{0}   (  x  )  )$ are obtained by using the highly accurate solitary wave solutions of the complete nonlinear water wave problem, provided by \cite{Ref71}. HCMS is implemented by using $N_{tot}     =    8$ modes and a spatio-temporal discretization $\delta x    =    0.2$ m and $\delta t    =$ $0.03$ s. The parameter $\mu _{{\kern 1pt} {\kern 1pt} 0} $ is chosen as $\mu _{{\kern 1pt}   0}     =    k\tanh   (  k  h_{  0}   )$ with $k    =    2\pi /L$, $L$ being the support of the free surface elevation as computed by the code of \cite{Ref71}.
 
\begin{figure*}[h]
 	\begin{center}
 		\includegraphics*[scale = 0.8]{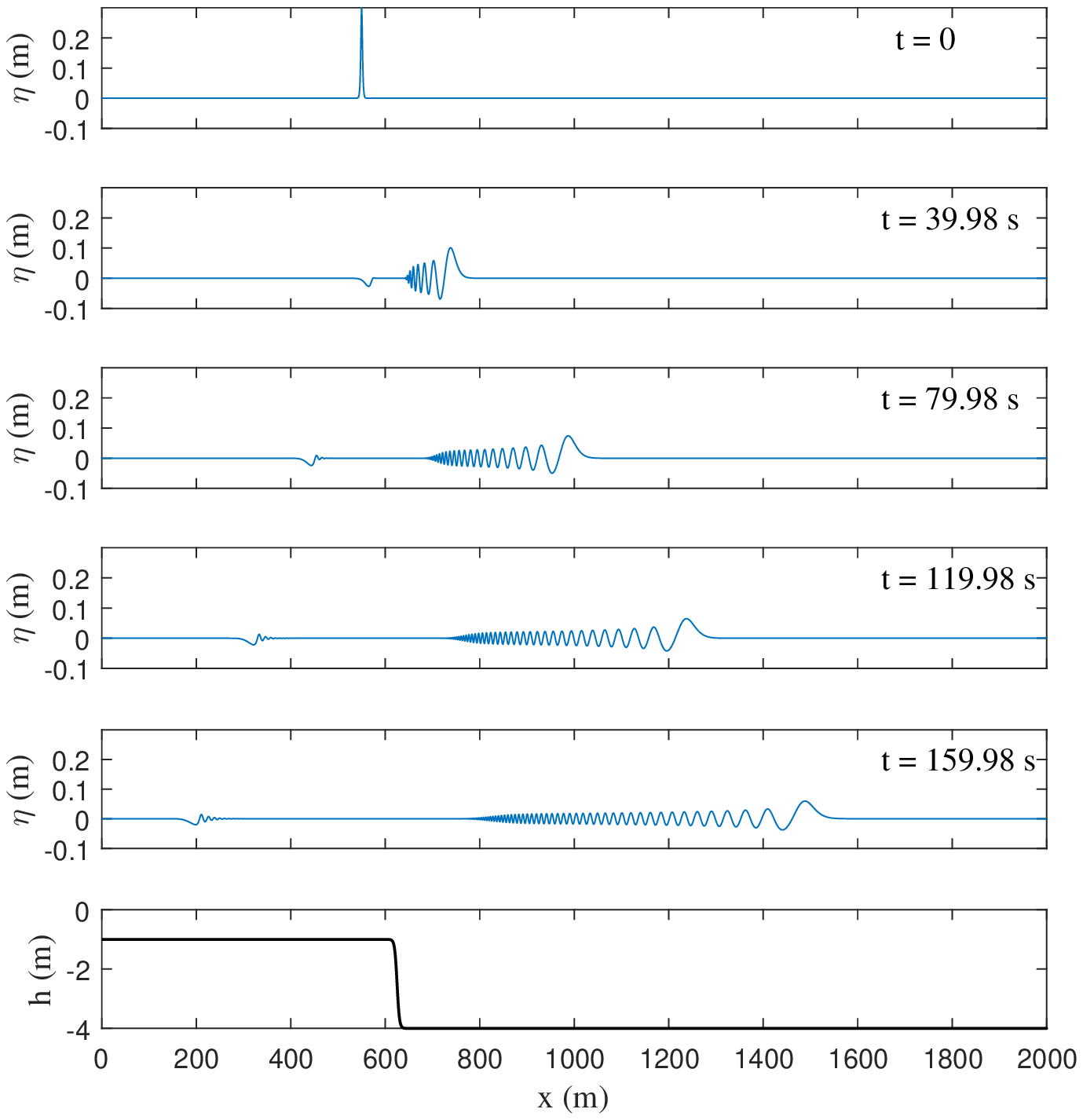}
 		\caption{ Snapshots of the free surface elevation for the simulation of a solitary wave propagating over an abrupt deepening, with $h_{  r}     =    4  m$}
 		\label{fig:17}       
 	\end{center}
 \end{figure*}

 In the first few seconds of the simulation, as the solitary wave is moving over the shallow-flat part of the bottom, it travels steadily towards the right. The moment it encounters the depending, a violent change of shape takes place. The crest height of the solitary wave drops suddenly, and a small wave of depression emerges, back-propagating towards the left. As the leading solitary-like wave travels towards the right, moving over the deep part of the seabed ($h_{  r}     =    4    m$), its crest height steadily reduces and a highly oscillating wave trail, of smaller amplitude, is being developed and expanded horizontally. This trailing wave follows the leading wave, giving rise to a dispersive wave pattern that propagates over intermediate-to-deep water conditions. Besides, as the reflected wave propagates towards the left, also develops a small yet visible dispersive trail. Some snapshots of the above described wave-bottom interaction pattern are shown in Figure \ref{fig:17}. The full simulation can be found in \href{https://www.dropbox.com/s/ijdqi8jr4y2q1f2/Solitary_Deepening_a03_4h_2000m_8modes.mp4?dl=0}{video 1}.

\begin{figure*}
 	\begin{center}
 		\includegraphics*[scale=0.7]{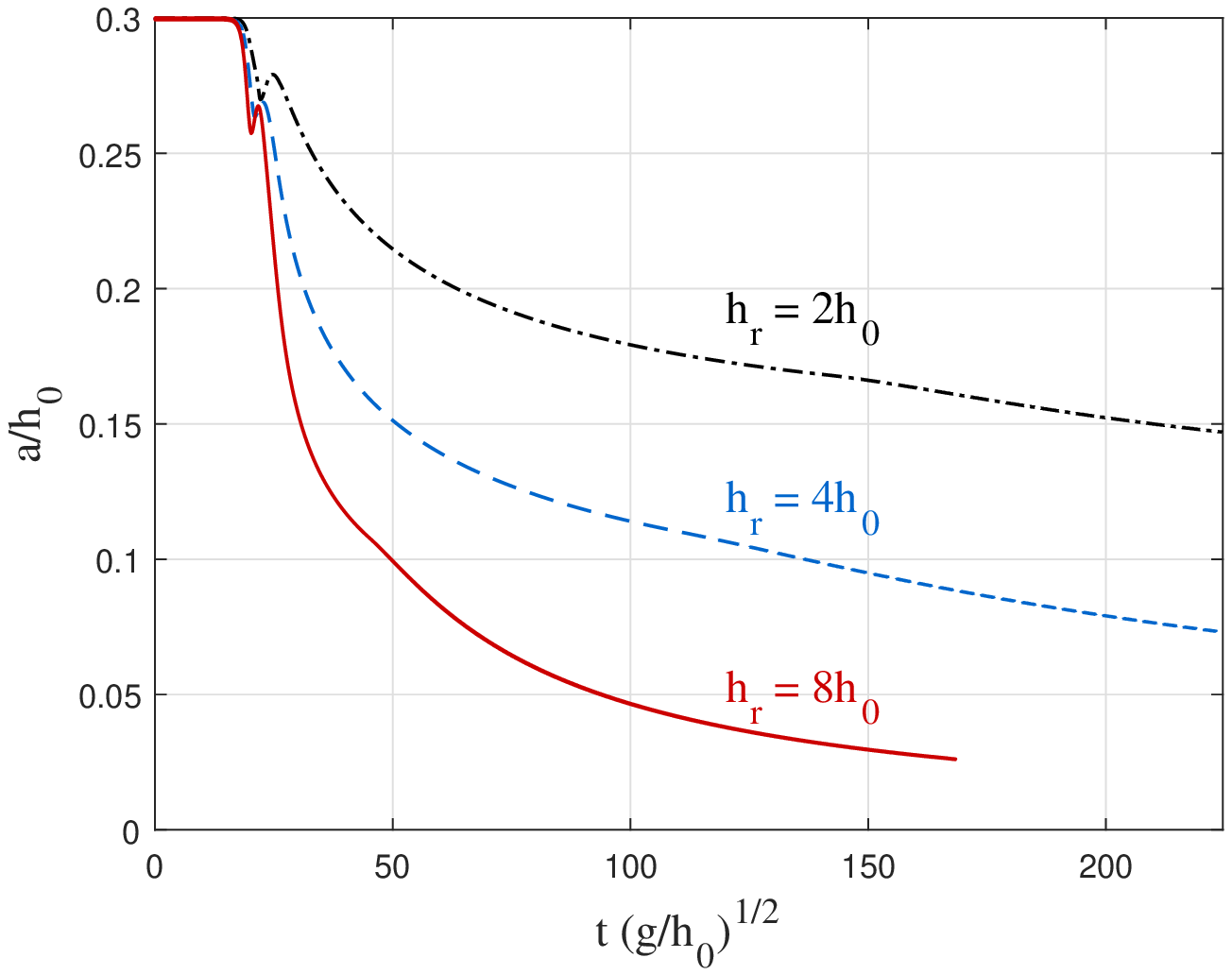}
 		\caption{ Maximum free surface elevation of the wave system propagating after the solitary wave passes over an abrupt deepening, from $h_{  0}     =    1  m$ to $h_{  r}     =    2  h_{  0}   {\kern 1pt} ,\,\,      4  h_{  0}   {\kern 1pt} ,\,\,  8  h_{  0} $}
 		\label{fig:18}       
 	\end{center}
 \end{figure*}

 In order to study the influence of the intensity of the deepening on the phenomenon, two further simulations are performed, with $h_{  r}     =    2  h_{  0} $ and $h_{  r}     =    8  h_{  0} $, keeping all other parameters the same. The time history of the maximum surface elevation along the evolution is shown in Figure \ref{fig:18}. In all cases, the initial sudden decrease of the amplitude is followed by a small and brief bounce after which the amplitude continues to decrease. The larger $h_{  r} $ the more rapid and significant the amplitude decrease. A long time after the interaction of the solitary wave with the abrupt deepening, the maximum elevation of the free surface continues to decrease (almost linearly) with time, at a slow rate. We believe that this phenomenon deserves a thorough experimental investigation, and we hope that it will be undertaken in the future.

 \section{Discussion and conclusions }
 \label{sec:7}

 The novel Hamiltonian Coupled-Mode Theory, proposed by \cite{Ref1}, \cite{Ref16}, is a non-perturbative, coupled-mode approach able to solve fully nonlinear water-wave problems in one or two horizontal dimensions, over varying bathymetry. In this approach, the evolution of the nonlocal Hamiltonian system requires the consecutive solution of a linear coupled-mode system with $(  \xh  ,  {\kern 1pt} t  )$-varying coefficients. The efficient implementation of the whole numerical scheme heavily relies on the fast and accurate evaluation of these coefficients, which are expressed in terms of the roots of the water-wave dispersion relation with $(  \xh  ,  {\kern 1pt} t  )$-varying frequency parameter.

 In this paper, we propose new, highly accurate, (semi)explicit formulae for the evaluation of wave numbers corresponding to evanescent modes. Their derivation is based on the successive application of a low-order iteration procedure (Picard iteration) and higher order methods, based on the general Householder formula for the solution of nonlinear algebraic equations. It is rigorously established that the compound iteration scheme, although retains the same order as that of the underlying high-order method, has a significantly lower error constant. The proposed 2${}^{nd}$ and 3${}^{rd}$ order semi-explicit methods are fairly simple and their first iteration provides quite accurate closed-form expressions for all eigenvalues $k_{  n} $ and all values of the frequency parameter $\mu $. Notwithstanding the usefulness of these expressions in all multimodal equations, their impact is really highlighted in connection with the HCMT, for solving fully nonlinear water wave problems over general bathymetry. In order to obtain stable long-time simulations in such demanding cases, computations at machine precision accuracy are necessary, and they are achieved by the present methods with not more than three iterations.

 The main conclusions of the present paper are: \textbf{(i)} the 2${}^{nd}$ and 3${}^{rd}$ order methods provide highly accurate results for all shallowness conditions, in contrast with the Newton-Raphson method that diverges in the deep water case, if the initial guess in not judiciously chosen; \textbf{(ii)} the 2${}^{nd}$ order method outperforms the 3${}^{rd}$ order one, in terms of computational time; \textbf{(iii)} The HCMT is efficiently implemented by using the closed-form expressions for the coefficients of the kinematical problem, evaluated by the 2${}^{nd}$ order semi-explicit formulae for the local wavenumbers.

\section*{Acknowledgement}
This research has not been supported by any funding bodies. The authors would like to thank Mr. A. Charalampopoulos for his support in the numerical simulations.


\appendix
\section{Proof of proposition 1}
\label{sec:App}
The proof will be derived for the general two-step recursive formula (compound scheme)
\begin{align}
{}^{j+1} \hat{x} & = \varphi  ({}^{j} x )\label{eqA1}\\
{}^{j+1} x    &=    G  ({}^{j+1} \hat{x}  )    =    G  (  \varphi   ({}^{j} x  )  )\label{eqA2}
\end{align}
where Eq. \eqref{eqA1} is a first-order iteration scheme which, for an appropriate ${}^{0} x$, converges to a simple root $x = a$ and ${}^{j+1} x = G ({}^{j} x )$ is an iteration scheme of order $p\ge 2$, converging to the same simple root. The last assumption implies, using a Taylor expansion with respect to $x = a$, that
\begin{equation}
\left|{\kern 1pt} {}^{j+1} x    -    a  \right|      \le     \frac{1}{p!}     \left|  G^{  (p)} (a)  \right|    \left|{\kern 1pt} {}^{j} x    -    a  \right|^{  p}     +      O{\kern 1pt}   \left(  \left|{\kern 1pt} {}^{j} x    -    a  \right|^{    p    +    1} {\kern 1pt} \right)
\end{equation}
Assume further that the functions ${\kern 1pt} \varphi   ,  {\kern 1pt} G$ are sufficiently smooth, and the distances $\left|{\kern 1pt} {}^{0} x    -    a  \right|$, $\left|  \varphi   {\kern 1pt} ({}^{0} x  )    -    a  \right|$ are such that the employed schemes converge to $a$. Since $\varphi   (a)    =    G  (a)    =    a$ and the derivatives $G^{  (k)} (a)    =    0$ for all $k<p$, from Eqns. (A2), following the same procedure used for the derivation of (A3), results to
\begin{equation}
G  {\kern 1pt} (  \varphi   {\kern 1pt} (x)  )      =      a      +      \frac{1}{p!}     G^{  (p)} (a)    \left[{\kern 1pt} \varphi '(a){\kern 1pt} \right]^{  p} (x-a)^{  p}       +  O\left((x-a)^{  p    +    1} \right)
\end{equation}
Thus, since $G  (\varphi   (a)  )    =    a$, for any iteration performed via Eq. (A2) it is
\begin{equation}
\left|{\kern 1pt} {}^{j+1} x    -    a  \right|      \le       R_{G}   \left|  \varphi '(a)  \right|^{  p}   \left|{\kern 1pt} {}^{j} x    -    a  \right|^{  p}     +      O{\kern 1pt}   \left(  \left|{\kern 1pt} {}^{j} x    -    a  \right|^{    p    +    1} {\kern 1pt} \right)
\end{equation}
 where $R_{G} =  \left|  G^{  (p)} (a){\kern 1pt} \right|/p!$. In the particular case of Picard iteration defined through Eq. \eqref{GrindEQ__21_}, we have $\varphi   (x)    =    n  \pi     -$ ${\rm Arctan}  {\kern 1pt} (\mu /x)$, hence
\begin{equation}
\left|  \varphi '  (\kappa _{n} )  \right|      =      \frac{\mu }{\mu ^{2} +    \kappa _{  n}^{  2} }       =      \frac{\mu }{\mu ^{2} +    (n  \pi     -    \varepsilon _{n} {\kern 1pt} )^{  2} }
\end{equation}
 From the definition of $\varepsilon _{n} $ it is $\varepsilon _{n} \in (  0  ,  {\kern 1pt} {\kern 1pt} \pi /2  )$ and thus
\begin{equation}
\left|  \varphi '  (\kappa _{n} )  \right|      \le       \frac{\mu }{\mu ^{2} +    (n  \pi -\pi /2)^{2} }       =      \frac{4\mu }{(2n-1)^{2} \pi ^{2} +    4\mu ^{2} }
\end{equation}
which completes the proof.
\end{document}